\def\kms{\hbox{km s$^{-1}$}}

\def\hii{H{\sc ii}\,}

\def\cob{CO(J=2$\rightarrow$1)}
\def\coc{$^{13}$CO(J=2$\rightarrow$1)}
\def\cod{C$^{18}$O(J=2$\rightarrow$1)}
\def\hcn{HCN(J=3$\rightarrow$2)}
\def\msun{M$_{\odot}$\,}
\def\jyb{Jy beam$^{-1}$}
\def\mjyb{mJy beam$^{-1}$}

\def\cm2{cm$^{-2}$}
\def\cm3{cm$^{-3}$}
\def\x{$\times$}

\def\radec{\hbox{RA, Dec. (J2000)}}

\def\ngc{\rm NGC\,3503}
\def\sfo{\rm SFO\,62\,}

\def\sfof{\rm SFO\,62}

\def\gra{$^{\circ}$}

\documentclass[printer]{aa}   

\usepackage{natbib}
\usepackage{graphicx}
\usepackage{txfonts}
\begin{document}
\title{  Molecules, dust, and protostars in  NGC 3503 }

\author{N. U. Duronea\inst{1,3}
          \and J. Vasquez \inst{1,2}
          \and G. A. Romero \inst{2,5}
          \and C. E. Cappa\inst{1,2}
          \and R. Barb\'a \inst{4}
          \and L. Bronfman\inst{3}
           }

\institute{Instituto Argentino de Radioastronomia, CONICET, CCT-La Plata, 
 C.C.5., 1894, Villa Elisa, Argentina   \email{duronea@iar.unlp.edu.ar}\and Facultad de Ciencias Astron\'omicas y Geof\'isicas, Universidad Nacional de La Plata, Paseo del Bosque s/n, 1900 La Plata,  Argentina\and Departamento de Astronom\'ia, Universidad de Chile, Casilla 36-D, Santiago, Chile\and Departamento de F\'isica, Universidad de La Serena, Cisternas 1200 Norte, La Serena, Chile\and Dr. Gisela Romero passed away on   January 18th. 2014}

\date{Received October 2013 / February 2014}
 
 
\abstract
   {}
   { We are presenting here a follow-up study of the molecular gas and dust  in the environs of the star forming region \ngc. This study aims at dealing with the interaction of the \hii\ region \ngc\  with its parental molecular cloud, and also with the star formation in the region, that was possibly triggered by the expansion of the ionization front against the parental cloud.    }
   {To analyze the molecular gas we use \cob, \coc, \cod, and \hcn\ line data obtained with the on-the-fly technique from the APEX telescope. To study the distribution of the dust, we make use of unpublished images  at 870 $\mu$m from the ATLASGAL survey and IRAC-GLIMPSE archival images. We use public 2MASS and WISE data to search for infrared candidate YSOs in the region.
}
   {The new APEX observations allowed the substructure of  the molecular gas in the velocity range from $\sim$ $-$28 to $-$23 \kms\ to be  imaged in detail. The morphology of the molecular gas close to  the nebula, the location of the PDR, and the  shape of radio continuum emission suggest that the ionized gas is expanding against its parental cloud, and confirm the ``champagne flow'' scenario. We have identified several molecular clumps and determined some of their physical and dynamical properties  such as density, excitation temperature, mass, and line width. Clumps adjacent to the ionization front are expected to be  affected by the \hii\ region, unlike those that are distant to it. We have compared the physical properties of the two kind of clumps  to investigate how the molecular gas has been affected by the \hii\ region. Clumps adjacent to the ionization fronts of \ngc\ and/or   the bright rimmed cloud   \sfo\ have been heated and compressed by the ionized gas, but their line width is not different to those that are too distant to the ionization fronts.  We identified several candidate  YSOs in the region. Their spatial distribution suggests that stellar formation might have been boosted by the expansion of the nebula. We discard the ``collect and collapse'' scenario and propose  alternative mechanisms such as radiatively driven implosion on pre-existing molecular clumps or small-scale Jeans gravitational instabilities.  
}
    {}


\maketitle

\section{Introduction}

   
It is accepted that OB associations have an enormous impact on the state of their environs. The interstellar medium (ISM) surrounding OB stars is expected to be strongly modified and disturbed  by their intense ultraviolet (UV) radiation field ($h\nu>$ 13.6 eV). UV photons ionize the surrounding gas creating \hii\ regions and dissociate the molecular gas originating photodissociation regions (PDRs) \citep{ht97}. Further, the surrounding neutral gas (either atomic or molecular), is compressed by the expansion of the \hii\ region and/or the action of stellar winds.  The compression of the cloud  could enhance the stellar formation via ``radiative driven implosion'' process (RDI;  \citealt{lela94}) or even trigger it via ``collect and collapse'' process \citep{elm77}. Therefore, when massive stars form inside a molecular cloud it is expected that they dominate the state of the parental cloud and consequently the further stellar formation process within. Indeed, it has been shown that a large fraction of stars originates at the peripheries of \hii regions \citep{pom09,ro09,ca09,deh10,va12,deh12}. Having this in mind, it is instructive to study the molecular gas adjacent to \hii regions  since it  can provide important information on the interaction  between massive stars and their natal environments. Furthermore, a comparison between regions of the molecular cloud adjacent to an \hii region with other sites away from it may provide significant understanding into the physical properties of the molecular gas that may impact the formation of stars.

\begin{figure*}
\centering
\includegraphics[width=470pt]{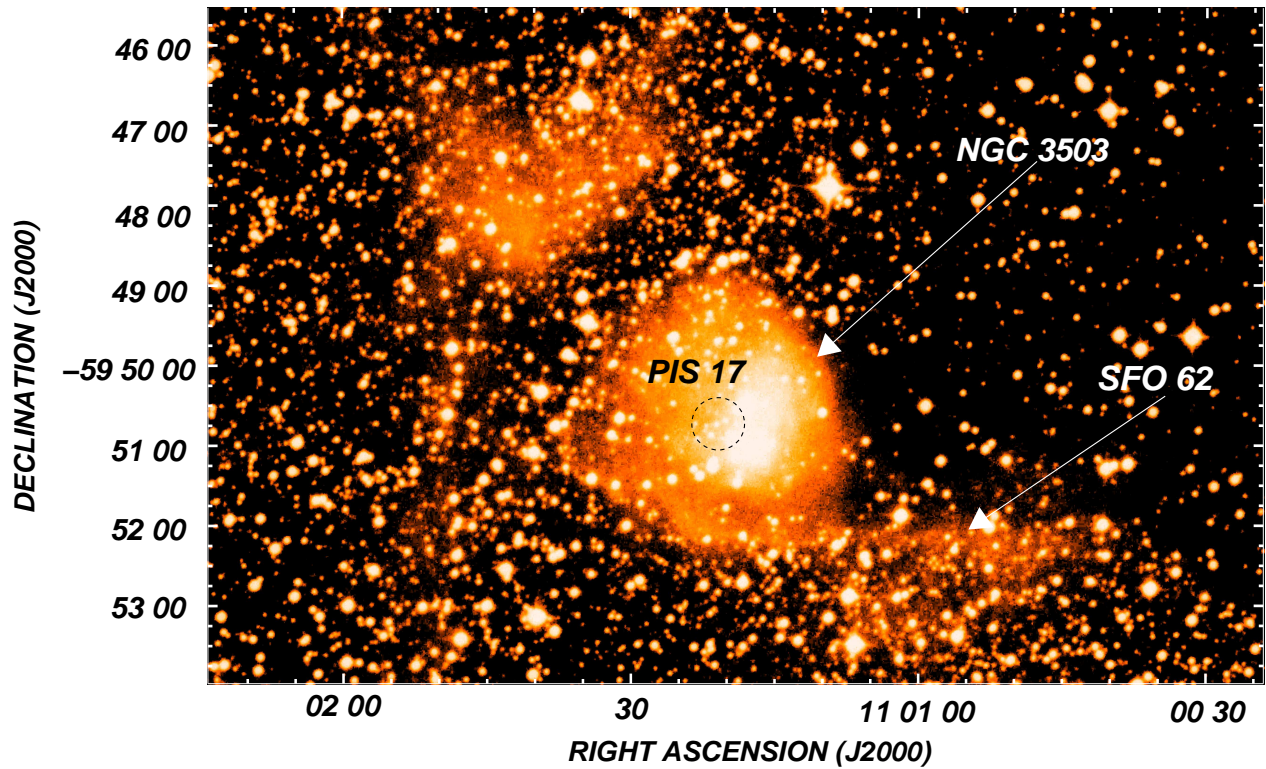}
\caption{UKST red plate image of the nebula \ngc\ in  the area covered by the APEX data. The positions of \ngc, the bright rimmed cloud \sfo, and the open cluster Pis 17 are indicated.}
\label{fig:ukst}
\end{figure*}

       The optical emission nebula NGC\,3503 (= Hf 44 = BBW 335) is a small \hii region located at  \radec\ = (11$^h$01$^m$16$^s$, $-$59\gra50\arcmin39\arcsec) \citep{dr88}, and placed at a distance of \hbox{2.9 $\pm$ 0.4 kpc} \citep{pco10}. \ngc\  is  ionized by early B-type stars belonging to the open cluster  Pis 17 \citep{H75,pco10,dvcca12} and is believed to be related to the bright-rimmed cloud (BRC)  {\rm SFO}\,62 \citep{so91,Y99,T04,U09}, although the ionizing star/s of the BRC have not  been certainly identified.  In a recent work (\citealt{dvcca12}; hereafter Paper I) we carried out a  multifrequency analysis in the environs of \ngc. We analyzed the properties of the molecular gas using NANTEN $^{12}$CO (J=1$\rightarrow$0)   (HPBW = 2.$'$7) observations, whilst the ionized gas was studied making use of radio continuum observations at 4800  and 8640 MHz,  with synthesized beams of \hbox{23\farcs 55 $\times$ 18\farcs 62}   and  \hbox{14\farcs 73 $\times$ 11\farcs 74}, respectively, carried out with  ATCA. The molecular line observations  revealed a molecular gas component of \hbox{7.6 $\times$ 10$^3$ \msun}in mass having a mean radial velocity\footnote{Radial velocities in this paper are always referred to the local standard of rest ({\rm LSR})} of $-$24.7 \kms (in agreement with the velocity of the ionized gas of \ngc; \citealt{g00}),  that  is associated with the nebula and its surroundings.  We reported an overdensity   centered at   \radec\ = (11$^h$01$^m$02.48$^s$, $-$59\gra50\arcmin04.6\arcsec\  \hbox{({\it l,b} = 289\fdg47, +0\fdg12)} (clump A)  projected near the border of \ngc, which is physically related to the nebula. Radio continuum images suggest that  the highest electron density area of the \hii region (coincident with the ionization front)     is compressing  the densest part of the molecular overdensity,  while the low electron density region  is undergoing a champagne phase.   Three MSX compact  \hii region (C\hii) candidates were also reported lying at the inner border of the nebula, which confirm the mentioned scenario (see Fig. \ref{fig:ir-cont}). In spite of the strong evidences of interaction between \ngc\ and its molecular environment,  disparities in angular resolution  made it difficult a direct comparison  between molecular and radio continuum/IR images.  Furthermore, the low angular resolution of the CO data did not allow us to detect any substructure in the molecular gas associated with \ngc, and therefore a detailed analysis could not be done adequately. 

The analysis of the molecular gas and dust associated with \ngc\ presents an important opportunity to study the interactions between \hii regions with molecular clouds, and how the formation of massive stars (Pis 17) in the cloud may  affect further (or ongoing) star formation.

In this paper, we present new  \cob,  \coc, \cod,  and   \hcn\  observations  around the \hii\ region \ngc\  carried out with the 12-m APEX\footnote{APEX is a collaboration between the Max-Planck-Institut fur Radioastronomie, the European Southern Observatory, and the Onsala Space Observatory} telescope.  To account for the  properties of  the  dust  in the nebula and its surroundings, we use unpublished images  at 870 $\mu$m from the ATLASGAL survey and available IRAC-GLIMPSE images. The aim of this work is to investigate in detail the spatial distribution and physical characteristics of the molecular gas and dust  associated with \ngc, and to compare them with the different gas components observed with similar angular resolutions. This analysis will also allow us to identify regions of   dense molecular gas where star formation may be developing.  To look for signatures of star formation,  a new  discussion of YSO candidates in the surroundings of  \ngc\    using WISE and 2MASS data is also presented. In Fig. \ref{fig:ukst}   we show the UKST red image\footnote{http://www-wfau.roe.ac.uk/sss/index.html} of the area of the nebula NGC3503 surveyed by our molecular observations


\section{Observations}

\subsection{Molecular observations}

The molecular observations presented in this paper were made during October 2011, with the Atacama Pathfinder Experiment (APEX) 12-m telescope \citep{gu06} at Llano de Chajnantor (Chilean Andes). As frontend for the observations, we used the APEX-1 receiver of the  Swedish Heterodyne Facility Instrument (SHeFI;  \citealt{vas08}). The backend for all observations was the eXtended bandwidth Fast Fourier Transform Spectrometer2 (XFFTS2)  with a 2.5 GHz bandwidth divided into 32768 channels. The observed transitions and basic observational parameters are summarized in Table \ref{lines}. Calibration was done by the chopper-wheel technique, and the output intensity scale given by the system is $T_{\rm A}$, which represents the antenna temperature corrected for atmospheric attenuation.  The observed intensities were converted to the main-beam brightness temperature scale by $T_{\rm mb}$ = $T_{\rm A}$/$\eta_{\rm mb}$, where   $\eta_{\rm mb}$ is the main beam efficiency. For the SHeFI/APEX-1 receiver we adopt $\eta_{\rm mb}$ = 0.75.

 Observations were made using the on-the-fly (OTF) mode with two orthogonal scan directions along RA and Dec.(J2000) centered on \radec = (11$^h$01$^m$16$^s$, $-$59\gra50\arcmin39\arcsec). For the observations  we mapped a region of $\sim$ 15$'$ $\times$ 10$'$. The spectra were reduced using the CLASS90 programme of the IRAM's GILDAS software package\footnote{http://www.iram.fr/IRAMFR/GILDAS}.

\begin{table}
\begin{center}
\caption{Observational parameters for the observed transitions.   }
\label{lines}
\begin{tabular}{lcccr}
\hline
molecular transition  & Frequency &  Beam & Velocity & rms  \\
      &  (GHz)    &  ($''$)     &resolution& noise \\
      &       &     & (\kms) & (K)\\ 

\hline
\hline
\cob      & 230.538000         &  $\sim$27       &  0.099      &   $\sim$0.3 \\
\coc     &  220.398677         &  $\sim$28     &   0.104     &  $\sim$0.3  \\
\cod     &  219.560357         &  $\sim$28      & 0.150       &  $\sim$0.25 \\
\hcn     &  265.886180         &  $\sim$23    &  0.086      &  $\sim$0.4\\
\hline 
\end{tabular}
\end{center}
\end{table}

\subsection{Continuum dust observations}

In this work we use unpublished images of ATLASGAL (APEX Telescope Large Area Survey of the Galaxy) at 870 $\mu$m (345 GHz)  \citep{sch09}. This survey covers the inner Galactic plane, $l$ = 300\gra\ to 60\gra, $|b|$ $\leq$ 1.\gra5, with a rms noise in the range 0.05 - 0.07 \jyb. The calibration uncertainty in the final maps is about of 15$\%$. LABOCA, the Large Apex BOlometer CAmera used for these observations, is a 295-pixel bolometer array developed by the Max-Planck-Institut fur Radioastronomie \citep{sir07}. The beam size at 870 $\mu$m is 19.$''$2. The observations were reduced using the Bolometer array data Analysis package (BoA; \citealt{sch12}).

\begin{figure*}
\centering
\includegraphics[width=470pt]{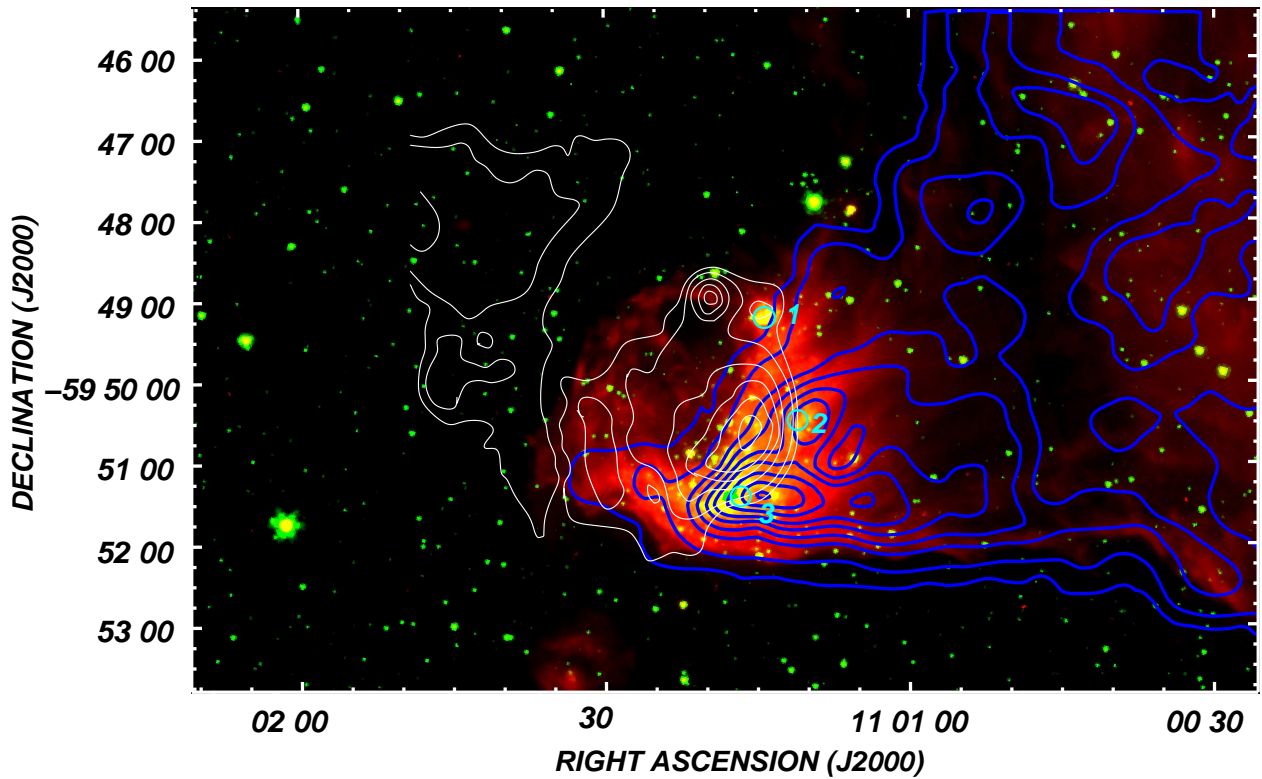}
\caption{ Composite image of \ngc\ and its environs. Red and green show emission at 8.0 and 4.5 $\mu$m (IRAC-GLIMPSE), respectively. White and  blue  contours show the radio continuum 4800 MHz and  integrated CO line emissions. Circles in light blue indicate the position of the three candidates to compact \hii\ region identified in the MSX catalog (see Table 3 in Paper I).   The CO temperature contours are 0.5  K \kms\  ($\sim$10 rms), 1.5, 6.6,  10.5, 15.5, 20.5,  25.5, and 30.5  K  \kms. The radio continuum 4800 MHz contours levels go from 2.4 \mjyb\ ($\sim$3 rms) to 10.4 \mjyb\ in steps of 2 \mjyb, and from 10.4 \mjyb   in steps of 4 \mjyb.         }
\label{fig:ir-cont}
\end{figure*}

\subsection{Physical parameters estimations}

\subsubsection{Excitation temperature and opacity}

 Excitation temperature ($T_{\rm exc}$) and opacity ($\tau$) of the \cob\ and \coc\ lines  in Local Thermodynamic Equilibrium (LTE) conditions,  can be derived using:
\begin{equation}
\label{eq:texc}
  T_{\rm peak} =\ T_0\ [J(T_{\rm exc})\ -\ J(T_{\rm bg})]\ \times\ [1\ -\ e^{(-\tau)}] 
\end{equation}
\citep{d78},  where   $T_0$ = $h \nu/ k$,  $J(T) = (e^{T_0 / T} -1)^{-1}$, and  $T_{\rm bg}$ is the background temperature (assumed to be $\sim$2.7 K).  The excitation temperature of the \coc\ line can be derived adopting  $T_{\rm exc}$($^{13}$CO) $\approx$ $T_{\rm exc}$(CO), which  can be calculated assuming that \hbox{$\tau^{12} >> 1$} and solving Eq. \ref{eq:texc} as follows:
\begin{equation}
\label{eq:texc2}
   T_{\rm exc}{\rm (CO)}\ =\ \frac{T_0^{12}}{{\rm ln} \left[ 1\ +\ \frac{T_0^{12}}{T_{\rm peak}{\rm  (CO)}\ +\ T_0^{12}\ \  [e^{(T_0^{12} / T_{\rm bg})} - 1]^{-1}     }\right]}\ 
\end{equation} 
 where $T_0^{12}$ = $h \nu^{12}/ k$, with $\nu^{12}$ = 230.538 GHz for the \cob\ line.  Then, the opacity  of the \coc\ line   can be derived solving again Eq. \ref{eq:texc}:
\begin{equation}
\ \  \tau^{13} = -{\rm ln}\ [1\ -\ (T_{\rm mb}^{13}/T_0^{13})\ [ (e^{T_0^{13} / T_{\rm exc}}\ -\ 1 )^{-1} -\  (e^{(T_0^{13} / T_{\rm bg})}-1)^{-1}    ]^{-1}  ], 
\label{tau13}
\end{equation}
 where $T_0^{13}$ = $h \nu^{13}/ k$, with $\nu^{13}$ = 220.399 GHz. We can also estimate the optical depth of the \cob\ line from the \coc\ line with  
\begin{equation}
\quad \tau^{12} =\  \left[\frac{\nu^{13}}{\nu^{12}}  \right]^2\  \left[\frac{\Delta {\rm v}^{13}} {\Delta   {\rm v}^{12}} \right]\ \left[\frac{\rm CO}{^{13}{\rm CO}} \right]\   \tau^{13},
\label{tau12}
\end{equation}
where   CO/$^{13}$CO is the isotope ratio (assumed to be $\sim$ 62; \citealt{lp93}).  $\Delta{\rm v}^{13}$ and $\Delta{\rm v}^{12}$ are defined as the {\it full width half  maximum} (FWHM) of the spectrum  of the $^{13}$CO and CO lines, respectively, which are derived by using a single Gaussian fitting  (FWHM = 2 $\times$ $\sqrt{2\ ln2}$ $\times$ $\sigma_{\rm gauss}$).

\subsubsection{Column density and mass}

Assuming LTE conditions  the H$_2$ column density, $N$(H$_2$), can be estimated from the  \coc\ line  following  the equations of  \citet{rw04}   
\begin{equation}
  N(^{13}{\rm CO}) = 1.5 \times 10^{14}\ \ \frac{ e^{[T_0(\nu_{10})/T_{\rm exc}]}\ \  T_{\rm exc} \int \tau^{13}\ \ d{\rm v}} {1 - e^{[ - T_0^{13}/T_{\rm exc}]}}   \quad \textrm{ (cm$^{-2}$)}
\label{n13co}
\end{equation}  
where $T_0(\nu_{10})$ = $h \nu_{10} / k$, being $\nu_{10}$   the frequency of the $^{13}$CO (1$\rightarrow$0) line (110.201 GHz).  To solve the integral of Eq. \ref{n13co} it can be assumed (in the limit of optically thin line) that $T_{\rm exc}\ \tau^{13}$ $\approx$ $T_{\rm mb}$.  In any case, optical depth effects can be diminished using the approximation
\begin{equation}
 T_{\rm exc}  \   \int{\tau^{13}} d{\rm v} \approx\ \ [\tau^{13}/1-e^{(-\tau^{13})}] \    \int{T_{\rm mb}}\ \ d{\rm v} 
\label{integral}
\end{equation} 
 This formula is accurate to 15 $\%$ for $\tau^{13} < 2$. We also estimate an uncertainty of $\sim$ 20 $\%$ from determining $\int{T_{\rm mb}}$ $d$v. The  molecular mass is then calculated using
\begin{equation}
  M(\rm H_2)\ =\  {\it m}_{\rm sun}^{-1}\ \ \mu\ \ {\it m}_H\ \ \sum\ \ \Omega\ \ {\it N}(\rm H_2)\ \ {\it d}^2 \quad  \quad \quad   \textrm{(M$_{\odot}$)}
\label{eq:masa}
\end{equation}
where  $m_{\rm sun}$ is the solar mass ($\sim$ 2 $\times$ 10$^{33}$ g),    $\mu$ is the mean molecular weight, which is  assumed to be equal to 2.72 after allowance of a relative helium abundance of 25\% by mass \citep{allen73},  $m_{\rm H}$ is the hydrogen atom mass   ($\sim$ 1.67 $\times$ 10$^{-24}$ g), $\Omega$ is the solid angle subtended by the CO feature  in ster, $d$ is the distance (assumed to be 2.9 $\pm$ 0.4 kpc; see Paper I) expressed in cm, and  $N$(H$_2$) is  obtained  using a ``canonical'' abundance \hbox{$N(\rm H_2)$ / $N(^{13} {\rm CO})$} = \hbox{5 $\times$ 10$^{5}$} \citep{d78}.

Considering only gravitational and internal pressure, neglecting support of magnetic fields or internal heating sources,  and  assuming a spherically symmetric cloud with a $r^{-2}$ density distribution, the virialized molecular mass, $M_{\rm vir}$, can  be estimated from
\begin{equation}\label{eq:virial}
\quad M_{\rm vir}\ =\ 126\ R_{\rm eff}\ (\Delta  {\rm v}_{\rm cld})^2  \qquad \qquad   \textrm{(M$_{\odot}$)}
\end{equation}
\citep{ml88}, where $R_{\rm eff}$ = $\sqrt{A_{\rm cloud}/ \pi}$ is the effective radius in parsecs,  and $\Delta {\rm v}_{\rm cld}$  is the width of the composite spectrum, defined as  for  $\Delta{\rm v}^{13}$ and $\Delta{\rm v}^{12}$ in Eq. \ref{tau12}. The composite spectrum is obtained by averaging all the spectra within the area of the cloud ($A_{\rm cloud}$). 

The mass of the dust can be calculated following  \citet{deh09}.   Considering that the emission detected at 870 $\mu$m originates in thermal dust  emission, the dust mass ($M_{\rm dust}$) can be derived  from
\begin{equation}\label{eq:mdust}
\quad   M_{\rm dust}\ =\  \frac{S_{870}\ d^2}{\kappa_{870}\ J_{870}(T_{\rm dust})}, 
\end{equation}
where $S_{870}$ is the measured flux density, $d$ is the adopted distance to the source, $\kappa_{870}$ is the dust opacity per unit mass at 870 $\mu$m , and $J_{870}(T_{\rm dust})$ is the Planck function for a temperature $T_{\rm dust}$.  We adopted a dust opacity $\kappa_{870}$ = 1.0 cm$^{2}$ g$^{-1}$ estimated for dust grains with thin ice mantles in cold clumps \citep{osse94}.

\section{Results and analysis of the observations}

\subsection{Spatial distribution of the molecular gas}

 To study the  molecular gas associated with \ngc\  we  have mostly focused on Component 1 of Paper I (which is   undoubtedly   associated with the nebula). We will also make a brief analysis of the molecular gas associated with Component 2 and Component 3. 

 In Fig. \ref{fig:ir-cont}     we show an overlay of the mean CO emission  as obtained with APEX, in the velocity interval from  $-$29 \kms\ to $-$24 \kms, onto the IRAC-GLIMPSE\footnote{http://sha.ipac.caltech.edu/applications/Spitzer/SHA/} \citep{ben03}     and 4800 MHz radio continuum emission\footnote{Obtained with ATCA (see Paper I)} of the nebula. Though a remarkable resemblance between CO  and IR emission (MSX-A band) was previously put forward in Paper I, specially towards the extended IR emission at north and east of \ngc\ (see Fig. 6 of that work), a tight morphological correlation between the IR nebula and the molecular gas in the studied velocity interval is revealed by our new APEX observations, which confirms that this cloud is physically associated with the IR nebula. From Fig. \ref{fig:ir-cont}, a  direct comparison of the molecular and 4800 MHz radio continuum emissions strongly suggests that the molecular gas is being compressed by an ionization front and is interacting with the nebula. On the other hand, the ionized gas seems to be expanding freely towards the opposite direction (i.e. the intercloud medium). The location of the PDR is traced by the emission of PAHs molecules in the 8.0 $\mu$m image (red colour). Since these complex molecules are destroyed inside  the ionized gas of an \hii\ region  (see \citealt{deh10} and references therein), they delineate the boundaries of \ngc.  The molecular gas clearly depicts the position of the ionization front.  As predicted in Paper I, no molecular emission is  detected in this velocity range towards the low electron density region, which  is compatible with  a champagne-flow scenario. However, we speculate that the IR arc-like filament (from here onwards the ``IR arc'')   seen in the northeastern border of the nebula at \radec\ $\sim$ (11$^h$01$^m$30$^s$, $-$59\gra50\arcmin10\arcsec) and \radec\ $\sim$ (11$^h$01$^m$25$^s$, $-$59\gra49\arcmin00\arcsec),  might be still associated with small amounts of molecular gas at velocities between $\sim$ $-$17.1 to $-$15.3 \kms\ (see Sect. 3.2).

\begin{figure*}
\centering
\includegraphics[width=410pt]{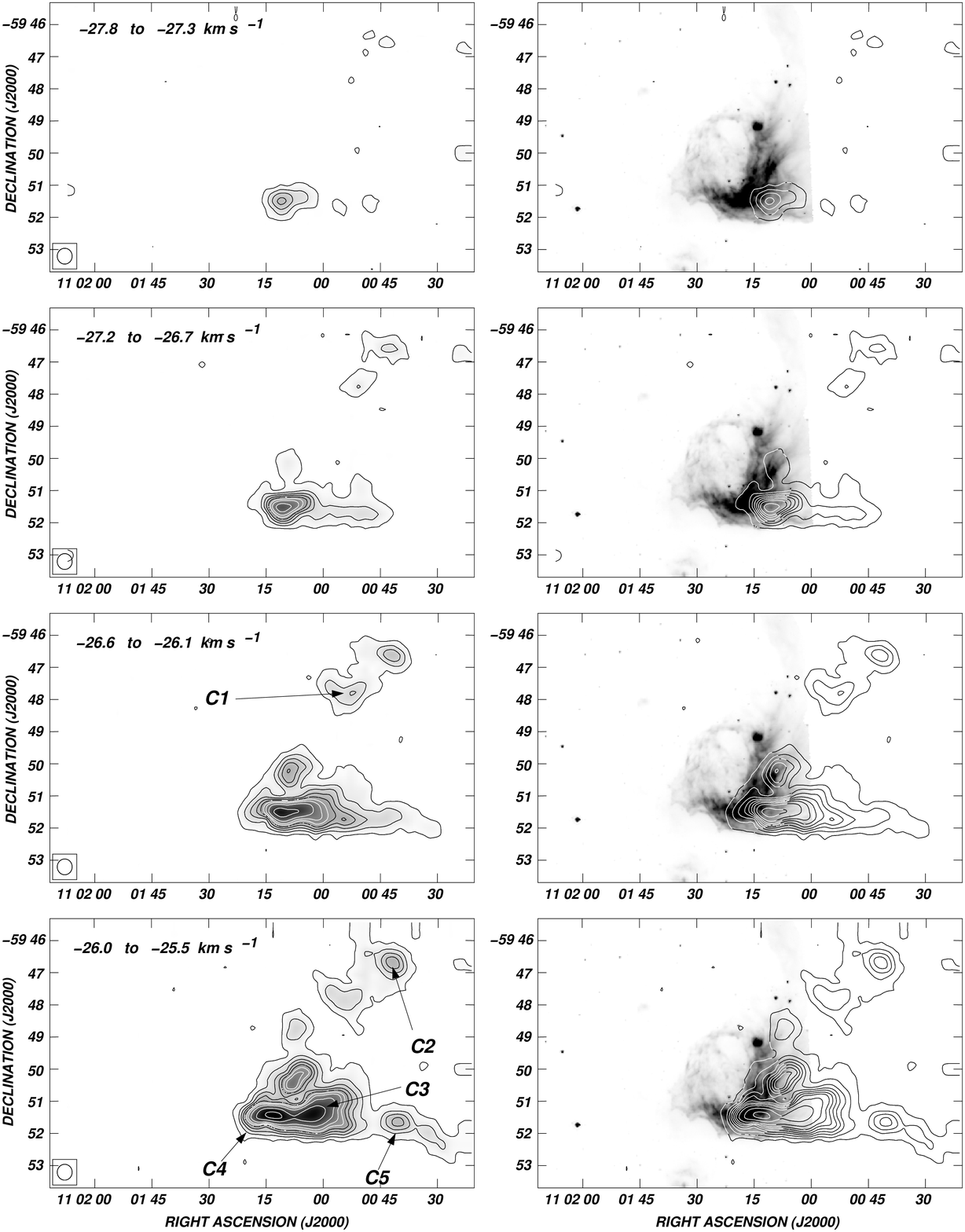}
\caption{{\it Left panels:}  Integrated $^{13}$CO emission in the velocity range from $\sim$ $-$27.8 to $-$23.7 \kms, and from 18.7 to 21.7 \kms\ (see continuation).   The velocity interval of each image is indicated in the upper left corner. The lowest temperature contour is 0.7  K \kms ($\sim$ 5 rms). The contour spacing temperature is 1.5  K \kms\ {\bf to} 9.7  K \kms, and 4  K \kms\ hereonwards.   The beam size is shown by a  circle  in the lower left corner of each image. {\it Right panels:} Overlay of the IRAC GLIMPSE 4 (8 $\mu$m)  emission of \ngc\ (grayscale) and the mean $^{13}$CO emission in the velocity intervals shown in the left panels (contour lines).    }
\label{fig:mosaico}
\end{figure*}

\begin{figure*}
\centering
\includegraphics[width=410pt]{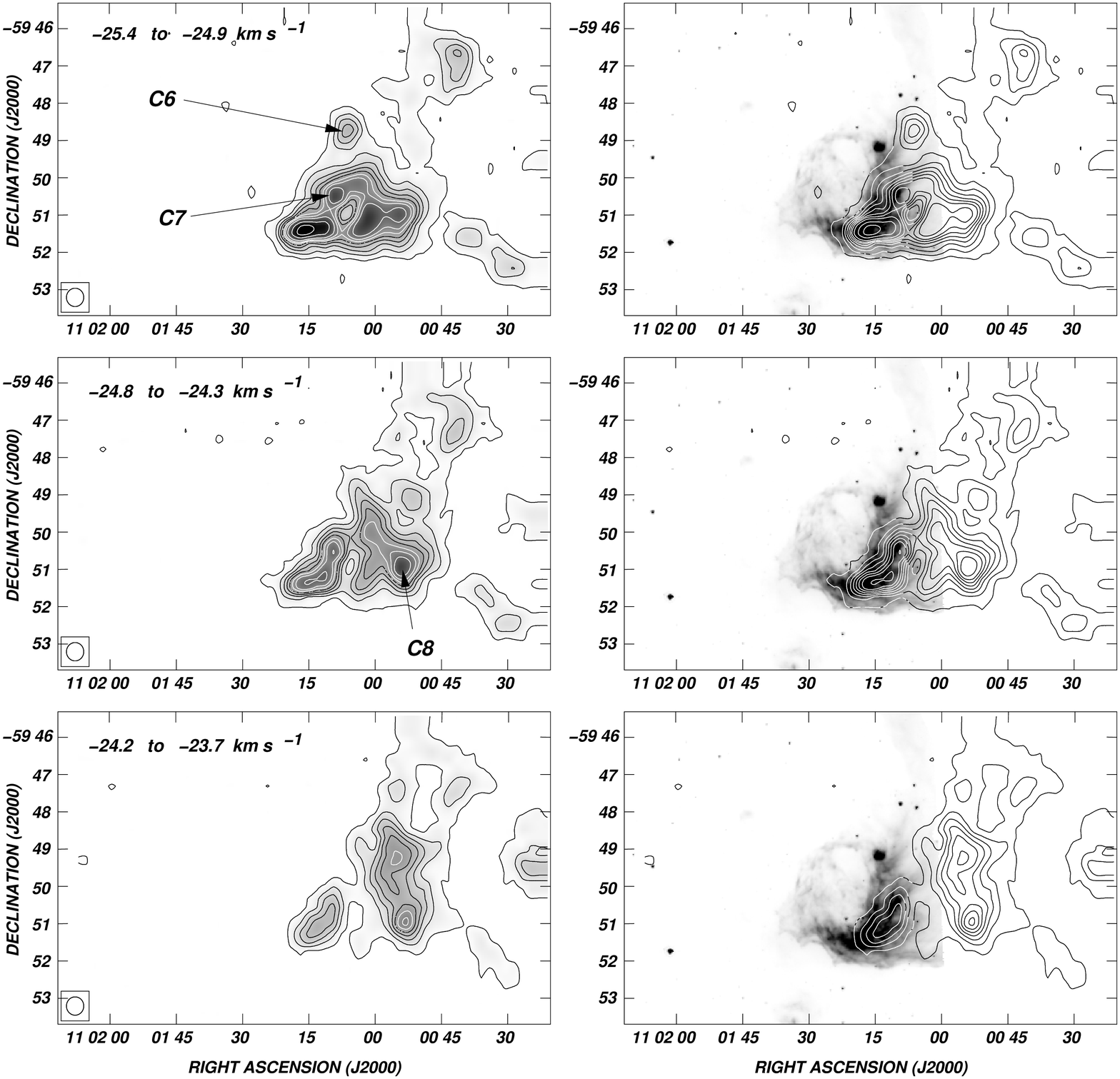}
\addtocounter{figure}{-1}
\caption{continuation}
\label{}
\end{figure*}

\begin{figure*}
\centering
\includegraphics[width=410pt]{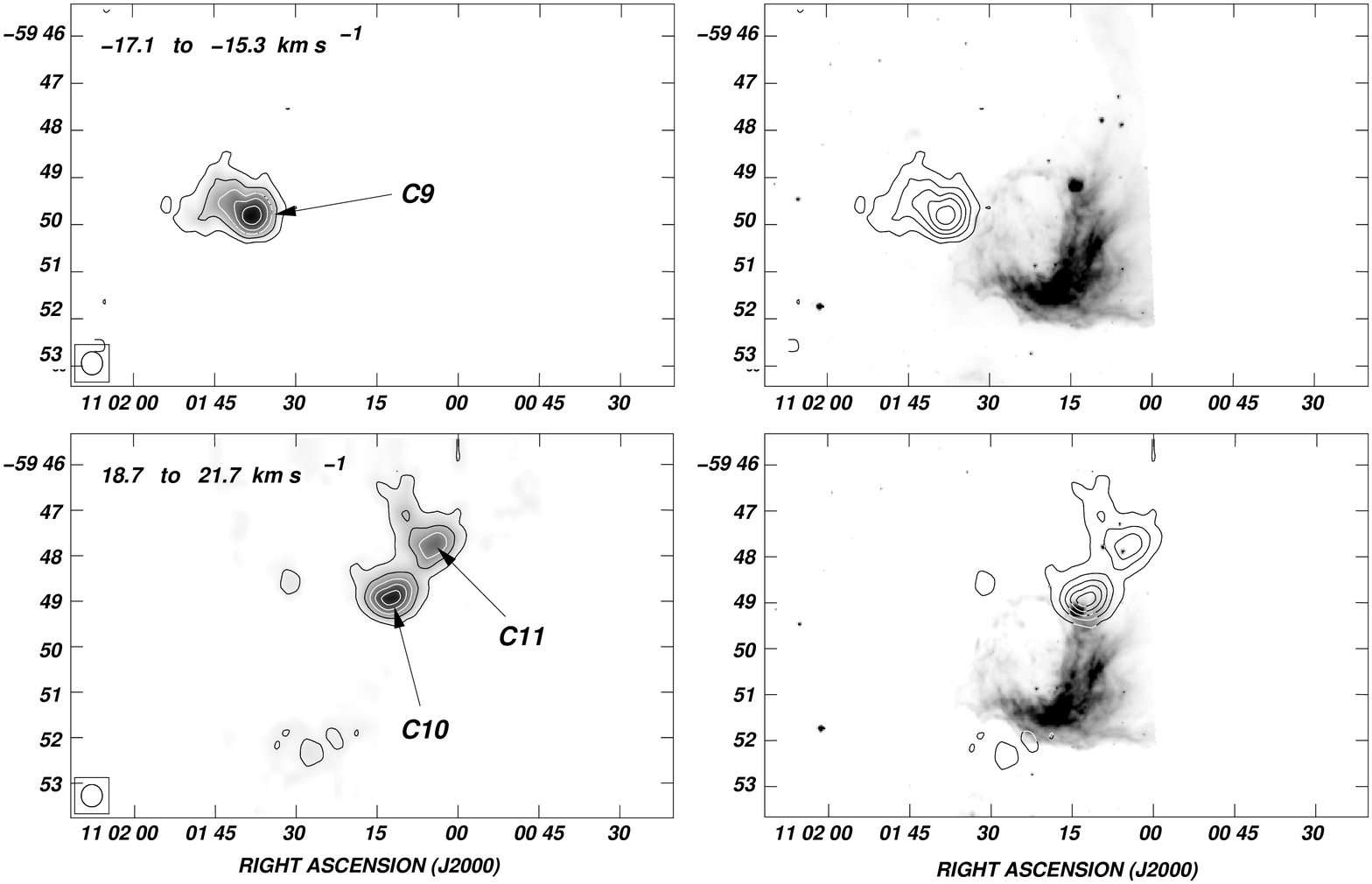}
\caption{{\it Left panels:}  Integrated $^{13}$CO emission in the velocity range from $\sim$ $-$17.1 to $-$15.3 \kms (upper panel), and from 18.7 to 21.7 \kms\ (lower panel).   The velocity interval of each image is indicated in the upper left corner. In the upper panel, the lowest $^{13}$CO temperature contour is 0.35  K \kms ($\sim$ 5 rms) and the contour spacing temperature is 0.7   K \kms.  In the lower panels, the  lowest  $^{13}$CO   temperature contour is 0.75  K \kms\ ($\sim$ 15 rms) and  contour spacing temperature is 1  K \kms. The beam size is shown by a  circle  in the lower left corner of each image. {\it Right panels:} Overlay of the IRAC GLIMPSE 4 (8 $\mu$m)  emission of \ngc\ (grayscale) and the mean $^{13}$CO emission in the velocity intervals shown in the left panels (contour lines).    }
\label{fig:c1y2}
\end{figure*}

Since the \cob\ line is optically thick and can be used only to trace low density gas, we analyze the $^{13}$CO data which allows one to go deeper into the molecular clouds, although it may fail to probe the densest molecular gas because it freezes onto dust grains at high densities \citep{mas07}. To study the kinematics of the molecular gas  in more detail  we have   averaged the individual original channel maps of the $^{13}$CO emission line. In Fig. \ref{fig:mosaico} (left panels) we show a collection of narrow velocity images depicting the mean  $^{13}$CO spatial distribution in the velocity  range from  $-$27.8   \kms\  to $-$23.7  \kms. Every image represents an  integral of the $^{13}$CO emission \hbox{$\int$ $T_{\rm mb}\ d{\rm v}$} \hbox{(K \kms)}  over a velocity interval of 0.5 \kms. In this way, the final rms noise for each interval results  $\Delta T_{\rm rms}$ $\approx$ 0.15 K \kms.     To analyze the relation between the molecular gas and the warm dust in the nebula, in the right panels of  Fig. \ref{fig:mosaico} we show the  $^{13}$CO emission in the velocity intervals mentioned before (in contours) projected onto the  IRAC GLIMPSE-4 (8 $\mu$m)  emission of \ngc\ (grayscale). In order to study the small-scale structure of the molecular cloud we have identified a number of small molecular clumps.     The clumps were selected by eye and based in the  following  simplest  criteria: 1) the peak temperature of each clump is at least 5 times the rms noise,  2)  the decrease in $T_{\rm mb}$ between the peak temperature of two adjacent clumps is larger than 5 times the rms noise, and 3)  the clump is present at least   along $\sim$40 $\%$ of the total velocity interval of the cloud.   We define the area of each clump ($A_{\rm clump}$) by the polygon that encloses the emission corresponding to half the $T_{\rm mb}$ peak   in the velocity interval at which the clump is observed.     The clumps  are indicated in Fig. \ref{fig:mosaico}  from C1 to C8 and  the labeling was made in     the velocity interval at which they reach the maximum emission peak temperature. Such molecular clumps are also identified in CO showing more extended emission.


The molecular emission becomes first noticeable in the  velocity interval from   $-$27.8 \kms\  to $-$27.3 \kms\   as a weak patchy structure centered at  (\radec) $\simeq$ (11$^h$01$^m$10$^s$, $-$59\gra51\arcmin30\arcsec). Within the velocity range from  $-$27.2 \kms to $-$26.7 \kms\  most of the emission comes from a broad and  cometary head-tailed structure  lying      along Dec.(J2000) $\approx$ $-$59\gra51\arcmin30\arcsec\ that is coincident with the optical feature \sfof\ (see Fig. \ref{fig:ukst}). This coincidence might confirm that this molecular feature is being ionized from a stellar source/s at lower declinations giving rise to the  BRC,  as previously suggested in Paper I. This structure shows a sharp cut-off in the direction of  \ngc\  which suggests an interaction between the \hii\ region and the molecular gas. Very likely, the molecular gas has undergone compression on the front side due to the expansion of the ionized gas and/or the stellar winds of members of Pis 17.  This trend is also observed in the CO emission.      Between  $-$26.6 \kms\  to $-$26.1 \kms\      a molecular emission maximum, detected close to the brightest IR region of \ngc\      at (\radec) $\approx$ (11$^h$01$^m$15$^s$, $-$59\gra51\arcmin25\arcsec), appears to be merged with the lengthened   molecular structure along Dec(J2000) $\approx$ $-$59\gra51\arcmin30\arcsec. A  molecular clump, C1, is indicated in this velocity interval, located at  (\radec) $\approx$ (11$^h$00$^m$50$^s$, $-$59\gra47\arcmin57\arcsec). This clump is detected in the total velocity range from $-$27.1 to $-$25.6 \kms.

In the velocity range from $-$26.0 \kms\  to $-$25.5 \kms\ the molecular emission shows a good morphological resemblance with the IR nebula. Four molecular clumps achieve the maximum emission temperature at this velocity range. Clump C2 is placed adjacent to C1, at  (\radec) $\approx$ (11$^h$00$^m$40$^s$, $-$59\gra46\arcmin40\arcsec) and is   identifiable in the total velocity range from $-$27.3 to $-$24.3 \kms.  A moderately intense extended emission seems to be connecting both clumps, which may indicate a physical association.  Clumps C3 and C4 are located at  (\radec) $\approx$ (11$^h$01$^m$05$^s$, $-$59\gra51\arcmin30\arcsec) and (\radec) $\approx$ (11$^h$01$^m$12$^s$, $-$59\gra51\arcmin25\arcsec), respectively. Their locations are coincident  with the cometary head-tailed feature observed between $-$27.2 \kms to $-$26.7 \kms which means that they all are part of the same molecular structure.  This correlation suggests that the formation of clumps C3 and C4 are the result of a fragmentation due to the compression of the \hii\ region  \citep{wi94}, although previous estimates seem to discredit this conjecture (see Paper I).       The molecular emission in the direction of these clumps is detected almost in the entire velocity range.  The position of C4 is highly coincident with an MSX compact \hii\ region candidate labeled  as source 3 in Paper I (see Fig. \ref{fig:ir-cont}). 
\begin{figure*}
\centering
\includegraphics[width=390pt]{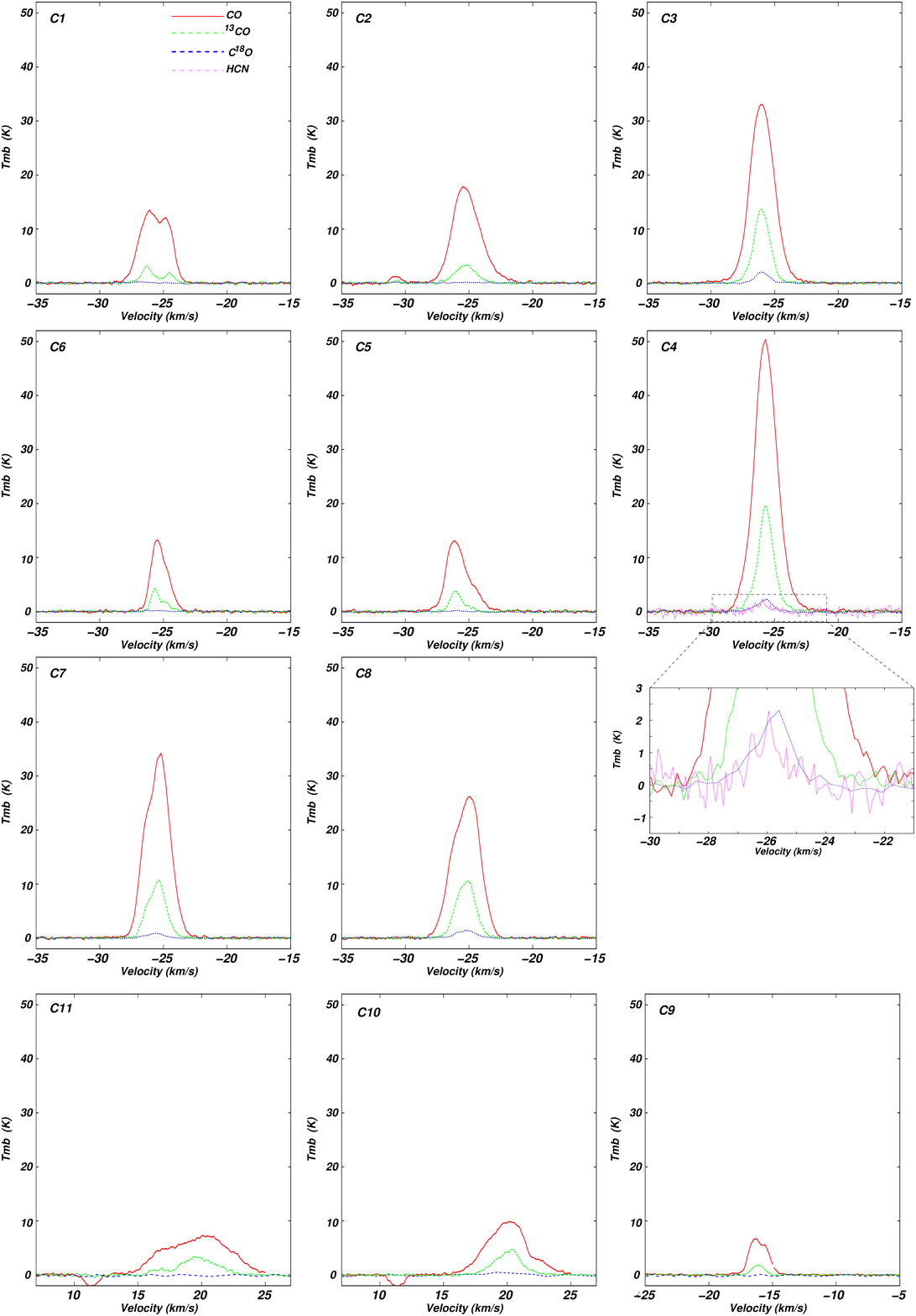}
\caption{ CO, $^{13}$CO, and C$^{18}$O  averaged line spectra obtained towards the molecular clumps C1 to C11. For C4, the spectrum of HCN is zoomed in the  middle right panel of the figure. }
\label{12-13-18}
\end{figure*}
This suggests that C4 is a high density molecular clump that is been irradiated by the UV field of \ngc.     The fourth molecular clump detected in this velocity range, C5, is located at (\radec) $\approx$ (11$^h$00$^m$40$^s$, $-$59\gra51\arcmin30\arcsec). Its location is almost adjacent with the westernmost border of \sfo\   (see Fig. \ref{fig:ukst}) and its eastern border seems to be connected with the western side of C3 by a weak bridge of $^{13}$CO emission. The molecular emission of C5  is detectable up to  $-$24.9 \kms. C3, C4, and C5 are detected in  the total velocity range from $-$27.1  to $-$24.9 \kms,  $-$27.8  to $-$24.5 \kms, and $-$26.6 \kms\ to $-$24.6 \kms, respectively. Probably, the leakage of UV radiation of \sfo\  into the molecular gas might have helped to shape clumps C3,  C4, and C5 (e.g. \citealt{pom09}).

\begin{table*}
\caption{Geometrical and line emission parameters obtained for the molecular clumps}
\centering
\begin{tabular}{lccccccccc}
\hline
Clump  &  RA               &  Dec.(J2000)             &  $A_{\rm clump}$($^{13}$CO)  &  $T^{13}_{\rm peak}$  & $T^{18}_{\rm peak}$ & $T^{12}_{\rm peak}$ & $T^{\rm HCN}_{\rm peak}$ & v$^{13}_{\rm centr}$ &  Vel. int. ($^{13}$CO) \\
       & ($^h$\ $^m$\ $^s$) & (\gra\ \arcmin\ \arcsec) &  (10$^{-8}$ ster)         &   (K)             &   (K)            &   (K)               & (K)    &  (\kms)   &     (\kms) \\ 
\hline
\hline
C1  &   11 : 00 : 52    &   $-$59 : 47 : 57       &  7.8       &     3.2            &   -       &  13.6   & -   & $-$26.3  &  $-$27.1\ to\ $-$25.6   \\
C2  &   11 : 00 : 40    &   $-$59 : 46 : 40       &  7.3       &     3.4            &   -       &  17.9   & -   & $-$25.3  &  $-$27.3\ to\ $-$24.3     \\
C3  &   11 : 01 : 05    &   $-$59 : 51 : 30       &  10.7      &     13.7           &  $\sim$ 2.1      &  33.2   & -   & $-$26.0  &  $-$27.1\ to\ $-$24,9    \\
C4  &   11 : 01 : 12    &   $-$59 : 51 : 25       &  8.6       &     19.7           &  $\sim$ 2.3      &  50.3   & $\sim$ 2.3 & $-$25.7  &   $-$27.8\ to\ $-$24.5   \\
C5  &   11 : 00 : 40    &   $-$59 : 51 : 30       &  5.5       &     3.8            &  -        &  13.3   & -   & $-$26.1  &  $-$26.6\ to\ $-$24.6     \\
C6  &   11 : 01 : 06    &   $-$59 : 48 : 40       &  4.1       &     4.3            &  -        &  13.4   & -   & $-$25.7   &  $-$26.1\ to\ $-$25.0     \\
C7  &   11 : 01 : 10    &   $-$59 : 50 : 28       &  8.9       &     10.7           &  $\sim$ 0.9      &  34.2   & -   & $-$25.5  &  $-$27.1\ to\ $-$24.7     \\
C8  &   11 : 00 : 53    &   $-$59 : 51 : 00       &  8.6       &     10.5           &  $\sim$ 1.5      &  26.2   & -   & $-$25.2  &  $-$25.6\ to\ $-$24.1     \\
C9  &   11 : 01 : 37    &   $-$59 : 50 : 00       &  9.1       &     1.8            &  -        &  6.7    & -   & $-$16.1  &  $-$17.1\ to $-$15.3            \\
C10 &    11 : 01 : 11    &   $-$59 : 48 : 57       &  3.9       &     4.7            &  -        &  9.9    & -   & 20.1  &  17.2\ to 22.5            \\
C11 &   11 : 01 : 05    &   $-$59 : 47 : 40       &  6.0       &     3.3            &  -        &  7.3    & -   & 20.9  &   15.3\ to 22.8           \\ 
\hline
\end{tabular}
\label{tabla:medidas}
\end{table*}

\begin{table*}
\caption{Physical and dynamical properties derived for the molecular clumps.    }
\begin{center}
\begin{tabular}{lccccccccccc}
\hline
 Clump  & $T_{\rm exc}$ & $\tau^{13}$ & $\tau^{12}$ &  $\Delta {\rm v}^{\rm 13}$ & $\Delta {\rm v}^{12}$ &  $\frac{\Delta {\rm v}^{\rm 13}}{\Delta {\rm v}_{\rm th}^{\rm 13}}$   &   $N(\rm H_2)^{\rm peak}_{\rm LTE}$ & $N(\rm H_2)^{\rm level}_{\rm LTE}$ & $M_{\rm LTE}$ & $M_{\rm vir}$ &  $\frac{M_{\rm LTE}}{M_{\rm vir}}$   \\
  &   (K)   &     &                        &   (\kms)   & (\kms)     &     &  (10$^{21}$ cm$^{-2}$)   &  (10$^{21}$ cm$^{-2}$)  &  (\msun) & (\msun) &   \\ 
\hline
\hline
C1   & 18.7   &  0.26    & 7.1   &   0.91  &  1.92  & 5.4   &   1.2 $\pm$ 0.2   &  0.8 $\pm$ 0.2  & 11 $\pm$ 5   & 36 $\pm$  12   & $\sim$  0.3 \\   
C2   & 23.1    & 0.21    & 9.0   &   1.82  &  2.41 & 9.6    &  3.3 $\pm$ 0.7    &  2.2 $\pm$ 0.4   &  29 $\pm$ 15  & 170 $\pm$  55   & $\sim$   0.2   \\   
C3   &  39.1    & 0.51    & 19.6   &   1.48  &  2.20  & 5.9  &   13.8 $\pm$ 2.7  & 9.8 $\pm$   1.9  &  191 $\pm$ 95   & 147 $\pm$  50    &  $\sim$  1.3  \\   
C4   &  57.2     & 0.48    & 17.6   &   1.39  &  2.13  & 4.5   &  27.1 $\pm$ 5.4  & 18.2 $\pm$ 3.6   & 282 $\pm$ 140  & 120 $\pm$  41   & $\sim$   2.4  \\   
C5   & 18.3     & 0.33    & 12.7   &   1.03  &  1.53 & 6.1   &   1.6 $\pm$ 0.3  & 1.2 $\pm$  0.2  & 12 $\pm$ 6  &   51 $\pm$   17  & $\sim$   0.2 \\   
C6   & 18.5     & 0.38    & 13.1   &   0.77  & 1.25   & 4.5   &  1.3 $\pm$ 0.3    & 0.9 $\pm$ 0.2  &  7 $\pm$ 3   & 22 $\pm$  8  & $\sim$   0.3   \\   
C7   &  39.7    & 0.37    & 14.2   &   1.53  & 2.26 & 6.1   &   10.0 $\pm$ 2.1   &  6.5 $\pm$ 1.3  &  169 $\pm$ 84     & 144 $\pm$  49 & $\sim$ 1.2  \\   
C8   &  31.6    & 0.50   & 19.0   &  1.61   & 2.42   & 8.1  &   8.5 $\pm$ 1.7    &  5.8 $\pm$ 1.2  & 92 $\pm$ 46   & 165 $\pm$   56 & $\sim$   0.6   \\   
C9  &   11.5   & 0.29  & 10.3    & 1.15      &   1.81  &  8.3   &  1.4 $\pm$ 0.3    &  0.4 $\pm$ 0.1    & 6 $\pm$ 3     & 79 $\pm$ 28   & $\sim$ 0.1  \\ 
C10 &   15.2    &   0.62 & 18.5    &  1.92   & 3.62   &  12.4 &    8.0 $\pm$ 1.6    & 5.6 $\pm$ 1.1     &  303 $\pm$  90$^{(\dag)}$    & 408 $\pm$  83$^{(\dag)}$   &  $\sim$ 0.7 \\ 
C11 &   12.2   & 0.58   &  24.2   &  3.11   &  4.23  &  22.2 &    7.1 $\pm$  1.4  & 4.9 $\pm$ 1.0   & 408 $\pm$  82$^{(\dag)}$    & 1340 $\pm$  270$^{(\dag)}$  & $\sim$ 0.3  \\ 
\hline
\end{tabular}
\label{tabla:propiedades}
\end{center}
{\scriptsize {\bf Notes:} $(\dag)$ Values obtained adopting a kinematical distance of 8 kpc.  Uncertainties were calculated taking into account only innacuracies in the boundary selection. }
\end{table*}

In the velocity range from $-$25.4 to $-$24.9 \kms\ clumps C6  and C7 achieve their maximum emission temperature. Clump C6, located at  (\radec) $\approx$ (11$^h$01$^m$07$^s$, $-$59\gra48\arcmin40\arcsec),  is observed in a small velocity interval ($-$26.1 to $-$25.0 \kms) and is not morphologically correlated with the nebula. On the other hand, clump C7, placed at (\radec) $\approx$ (11$^h$01$^m$10$^s$, $-$59\gra50\arcmin30\arcsec), is projected onto the IR nebula and close to the location of the MSX compact \hii region candidate     labeled as source 2 in Paper I (see Fig. \ref{fig:ir-cont}). Clump C7 is detected over a velocity interval from $-$27.1 to $-$24.7 \kms.

In the velocity interval from $-$25.4  to $-$24.9 \kms clump C8 becomes noticeable as a patchy structure at (\radec) $\approx$ (11$^h$00$^m$52$^s$, $-$59\gra51\arcmin00\arcsec),    merged to clump C3. It achieves its maximum emission temperature in the velocity range   from $-$24.8  to $-$24.3 \kms. It is barely detected  beyond a velocity of $\sim$ $-$23.7 \kms.   Clump  C8  is almost adjacent to  SFO 62 which indicates that this clump might be affected by the BRC. 

It is worth mentioning that the \cod\ line emission, which is a good tracer of high density molecular gas, is only detected in the velocity range from $\sim$ $-$26.5 to $-$25.1 \kms, and is mostly concentrated towards the position of clumps C3, C4, and C7 (see Sect. 3.6)

  In Fig. \ref{fig:c1y2} (upper panels) we show the $^{13}$CO emission distribution integrated in the velocity interval from  $-$17.1 to $-$15.3 \kms. A relatively extended patchy structure is observed  at  (\radec) $\simeq$ (11$^h$01$^m$37$^s$, $-$59\gra50\arcmin00\arcsec).  For the sake of the convention, this structure is identified as clump C9.  This molecular clump  was reported in Paper I as Component 2. From a comparison with the IR emission (right panel) it can be noticed a good correspondence between the external border of C9 with the  IR arc which suggests a physical association between this clump and the nebula.

From a spectrum obtained in the direction of the MSX C\hii region candidate source 1  (from Paper I) we can observe that the  bulk of the molecular emission is in the velocity range from $\sim$ 19 to 22 \kms. In Fig. \ref{fig:c1y2} (lower panels)  we show the molecular emission integrated in the velocity range from 18.7  to 21.7 \kms. From this figure, it can be noted a molecular structure (from here onwards clump C10) almost projected over source 1. The circular galactic rotation model by \citet{bb93} locates this clump at 8 kpc. Probably, it is associated with the larger molecular structure reported in Paper I as Component 3. In the same velocity range another molecular structure (clump C11) is observed at (\radec) $\approx$ (11$^h$01$^m$05$^s$, $-$59\gra47\arcmin40\arcsec), which is probably associated with clump C10.   Since the radio continuum image at 4800 MHz indicates the presence of ionized gas in the  direction of C10, further radio recombination line observations may help to confirm or discard the velocity interval and a physical  association between this molecular clump  and source 1. Nevertheless, the disparity in the central velocities between C10 and C11 with the rest of the clumps ($\sim$45 \kms) suggests  that these clumps are physically unrelated to \ngc.  Assuming a distance of $\sim$ 8 kpc, clumps C10 and C11 turn out to be the most massive clumps in the sample (see Table \ref{tabla:propiedades}, Sect 3.2).

In Fig. \ref{12-13-18}  we show the  averaged \cob, \coc, and \cod\ spectra obtained inside the emission temperature level which defines each  $A_{\rm clump}$. In Table \ref{tabla:medidas} we present some morphological properties of the clumps and  averaged emission line parameters obtained in  their direction. Columns (2) and (3) give the coordinates of the center of the clumps.  Column (4) lists the area of each clump. In columns (5), (6), (7), and (8) the peak emission of the $^{13}$CO, C$^{18}$O, CO, and HCN lines are given. The C$^{18}$O and HCN peak temperatures  are listed only when  a value at least 3 rms noise is achieved.     In   column (9) we list the central velocity obtained by gaussian fitting  of the  integrated $^{13}$CO spectra within $A_{\rm clump}$, and column (10) indicates the  velocity interval of the $^{13}$CO line at which each clump is detected.

\subsection{Physical properties of the molecular gas}

In the previous section we analyzed the morphological and kinematical properties of the molecular gas in the velocity interval from $\sim$ $-$28 to $-$24 \kms,  $-$17.1 to $-$15.3 \kms, and 18.7 to 21.7 \kms\  in the environs of \ngc\ and \sfo.   Morphological characteristics  clearly evidence a physical association between the \hii\ region and the BRC with the molecular clumps identified within the  velocity interval from $\sim$ $-$28 to $-$24 \kms, and probably from $-$17.1 to $-$15.3 \kms. In this section, we analyze and compare the physical properties of the molecular clumps aimed at finding some influence of shock fronts or UV radiation on the molecular gas. We include a brief comment on C10 and C11 although, as mentioned before, a physical association of these clumps with \ngc\ is doubtful.

In Table \ref{tabla:propiedades} we list some important physical and dynamical properties derived for the molecular clumps using the  averaged spectra from Fig.  \ref{12-13-18}.  Column (2) lists the excitation temperature obtained from the CO peak and using Eq. \ref{eq:texc2}. Columns (3) and (4) give the optical depth of $^{13}$CO and CO, obtained with Eqs. \ref{tau13} and \ref{tau12}, respectively. The width of $^{13}$CO and CO lines ($\Delta {\rm v}^{\rm 13}$ and $\Delta {\rm v}^{\rm 12}$), defined as the FWHM of the line  (see Sect. 2.3.1), is tabulated in columns (5) and  (6), respectively. The ratio between $\Delta {\rm v}^{\rm 13}$ and the expected thermal width is listed in column (7). The expected thermal width of  $^{13}$CO is estimated using  $\Delta {\rm v}_{\rm th}^{\rm 13}  = \sqrt{8\ ln2\ k\ T_{\rm k} /  m}$, where $k$ is the Boltzman constant, $T_{\rm k}$ is the kinetic temperature (assumed to be equal to $T_{\rm exc}$), and $m$ is the mass of the  $^{13}$CO molecule.   The line widths  of the  CO and $^{13}$CO spectra  were derived from Gaussian fittings averaging all the spectra within $A_{\rm clump}$.  Columns (8) and (9) give the column densities at the emission peak and averaged within $A_{\rm clump}$, respectively. The mass derived assuming LTE and virial equilibrium is listed in columns (10) and (11), respectively.  There are some uncertainties in measuring $M_{\rm LTE}$ and $M_{\rm vir}$. In both cases, the values are affected by a distance indetermination of $\sim$ 15 $\%$ ($d$ = 2.9 $\pm$ 0.4 kpc) which yields to uncertainties of $\sim$ 30 $\%$ for $M_{\rm LTE}$ ($M_{\rm LTE}$ $\propto$ $d^2$)  and $\sim$ 15 $\%$ for $M_{\rm vir}$  ($M_{\rm vir}$ $\propto$ $d$).  In addition, inaccuracies in the boundary selection ($\sim$ 20 $\%$) can affect the size estimation of the clumps and therefore to produce significant uncertainties in the mass calculations, since a considerable part of the molecular mass could be missed,  especially for the cases of clumps C3, C4, C7 and C8 (they have very high emission temperature boundaries above the minimum 5 rms noise).  We estimate total uncertainties of about $\sim$ 50 $\%$ and $\sim$ 30 $\%$ for  $M_{\rm LTE}$ and $M_{\rm vir}$, respectively.  It is also worth mentioning that  when abundances and isotopologic ratios are considered,  accuracy    might be within a factor of 2. From Table \ref{tabla:propiedades}, it can be noted that $\Delta {\rm v}^{12}$ is, in average, larger than $\Delta {\rm v}^{13}$   by a factor of 1.5. Since the CO  molecule is optically thick,  some asymmetries in its spectra might be misjudged.  For example, clump C1 exhibits a double peak profile in its CO spectrum. This could be a result of self-absorption, which could indicate the existence of hot/warm gas inside the clump. While, the emission of $^{13}$CO clearly suggests the existence of a double cloud   in the line of sight at different velocities (see Figs. \ref{fig:mosaico} and  \ref{12-13-18}). Then, having the two isotopes provides a tool to discern the numbers of components along the line of sight. Similarly, the CO spectra of C5 and C6 show small ``shoulders''. An inspection at their $^{13}$CO spectra suggests that these shoulders  are the result of a second weaker component at more positive velocities (see also Fig. \ref{fig:mosaico}). As a result, the virial masses using   $\Delta {\rm v}^{12}$   might be overestimated. Then, we used   $\Delta {\rm v}^{13}$ for the calculations. For C1, C5, and C6 we take into account only the strongest molecular components (at more negative velocities).

Molecular clumps that are close to the ionized gas are expected to have different properties than those distant  from it, mostly due to  shock fronts and stellar FUV radiation  impinging  onto the cloud. To search  for their influence  on the molecular gas we  analyze the physical properties listed in Table \ref{tabla:propiedades}.

An analysis of temperatures and densities would be very instructive in this matter.  Since $T_{\rm exc}$ is derived using the optically thick \cob\ emission, it probes the surface conditions of the clouds. An inspection at Table \ref{tabla:propiedades} shows that clumps C1, C2, C5, and C6 (from here onwards the ``cold clumps'') achieve temperatures in a range between $\sim$ 18 - 23 K. These values are typical in cold molecular clouds, where cosmic ray ionization is  the main heating source.  Clumps C3, C4 and  C7 (from here onwards the ``warm clumps'') achieve temperatures in the range $\sim$ 39 - 57 K  which suggests that additional local heating sources are present. Clumps C4 and C7 lie at the edge of \ngc\ and they appear to have been  externally heated through the photoionization of their surface layers, as proposed in Paper I. This is reinforced by the presence of a PDR at the interface between the nebula and the clumps (see Fig. \ref{fig:ir-cont}). For the case of C3, its location (adjacent to \sfo, see Figs.  \ref{fig:ukst} and  \ref{fig:ir-cont}) would explain its high temperature (39.1 K). It is also worth to mention that the warm clumps  have also higher column densities, which very likely indicates that they  are actually formed by gas that has been swept up by the expansion of the ionization and has been condensed. An inspection at Figs. \ref{fig:ukst} and  \ref{fig:ir-cont} shows that C4 is ``trapped'' between two ionization fronts (\ngc\ and \sfo). Then, there might be additional compression, heating and ionization acting  upon C4, which might explain the high surface temperature and density derived for this clump. Furthermore, it is the only clump detected in HCN emission (see Sect. 3.4). We keep in mind, however, that Pis 17 has probably been formed inside high density molecular gas nearby to C4 and C7 that was later evacuated, so an increment in the density of the molecular environment is expected.  In addition, star forming process that are likely occurring inside clumps C3, C4 and probably C7 (see Sect. 3.6) may be contributing in rising the  temperature. Temperature and density of C8 are above than those of cold clumps,   which might implicate external sources of heating and compression. Although this clump appears to be more distant of \sfo\ than C3,  an interaction with the ionized gas of the BRC  cannot be discarded.

It is also  expected that clumps neighboring the  \hii\ region exhibit signs of turbulence  which could be manifested as line widths significantly broadened. However, an inspection of Table \ref{tabla:propiedades} shows that  all the clumps exhibit line broadenings beyond the natural thermal width. Different from  expected,  the molecular emission towards clump C2 (which is the most distant to an ionization front),  shows the broadest averaged spectra, while clump  C4, which clearly shows signs of interactions with the ionization fronts of \ngc\ and \sfo (see the above paragraph),  exhibits  the lowest ratio between observed and thermal widths.  Probably, the line width of the averaged spectrum reflects that different parts of the clouds inside $A_{\rm clump}$ have different velocities, rather than showing turbulence effects. Further, the angular resolution of the observations may not suffice to resolve objects and/or effects in the molecular gas (e.g. outflows, infall, accretion, etc.) that may be contributing in the broadening of the emission line.  

 Warm clumps have  the highest  LTE masses (71 to 282 \msun), in comparison with  colder clumps (7 to 29 $M_{\odot}$). This trend is also  observed with virialized masses with the exception of C2. An inspection at Table \ref{tabla:propiedades} shows that for the case of  cold clumps the ratio $M_{\rm LTE}/M_{\rm vir}$ is in the range 0.2 to 0.3.  On the other hand, for the case of warm clumps mass ratios $M_{\rm LTE}$/$M_{\rm vir}$ $>$ 0.6 were obtained (predominantly $>$1). Special attention may require for the case of clump C4, since its $M_{\rm LTE}$ is more than  twice larger than $M_{\rm vir}$.  Classical Virial equilibrium analysis establishes that the virial mass is the minimum mass required in order for a cloud to be gravitationally bound. Then, if $M_{\rm vir}$    is larger than $M_{\rm LTE}$   the cloud has too much kinetic energy  and is unstable. That seems be the case for cold clumps which are probably  expanding as a result of a lack of an external stabilizing    pressure. Differently,  warm clumps seem to be in virial equilibrium (gravitationally bound) which suggests that the ionization front and UV radiation of the \hii\ region is sufficient to heat up, but not enough to disrupt the molecular gas of the clump.   The presence of two protostellar candidates projected onto the direction of C4 (see Sect. 3.6) might indicate that infalling motions may be occurring inside this clump, which is in line with its high  $M_{\rm LTE}$/$M_{\rm vir}$ ratio.  Nevertheless, new observational results suggest that physical properties of molecular clouds do not agree with the classical interpretation of virial equilibrium (balance between gravitational and kinetic energies)  in the sense that an external pressure acting like confining force is needed \citep{hey09,fie11}.        We also keep in mind the caveats in the line width determination (see previous paragraph), and hence in determining $M_{\rm vir}$,   which could lead us to mistakenly interpret the results. Higher spatial resolution  observations might help to achieve  more conclusive results.

 We have excluded clumps C9, C10, and C11 from the previous analysis, since it is not clear whether they are physically associated with \ngc.  As mentioned in Sect. 3.1, the external border of C9 shows a good correspondence with  the IR arc. The low density and mass  derived for C9 (see Table \ref{tabla:propiedades}) suggests that  this feature might be some residuary molecular gas from  the original parental cloud after the \hii\ region have blown into the intercloud medium. The location of C9 (in the opposite side of clumps C4 and C7) and its velocity (shifted by $\sim$ 10 \kms) is in line with this scenario. Clumps C10 and C11 display the largest line widths (and  low excitation temperatures), with complex spectra typical of the presence of several velocity components. These characteristics add more support to our previous assumption that clumps C10 and C11 are physically unrelated to the rest of the molecular gas associated with the nebula.

\subsection{The velocity field of C4}

\begin{figure}[h!]
\centering
\includegraphics[width=240pt]{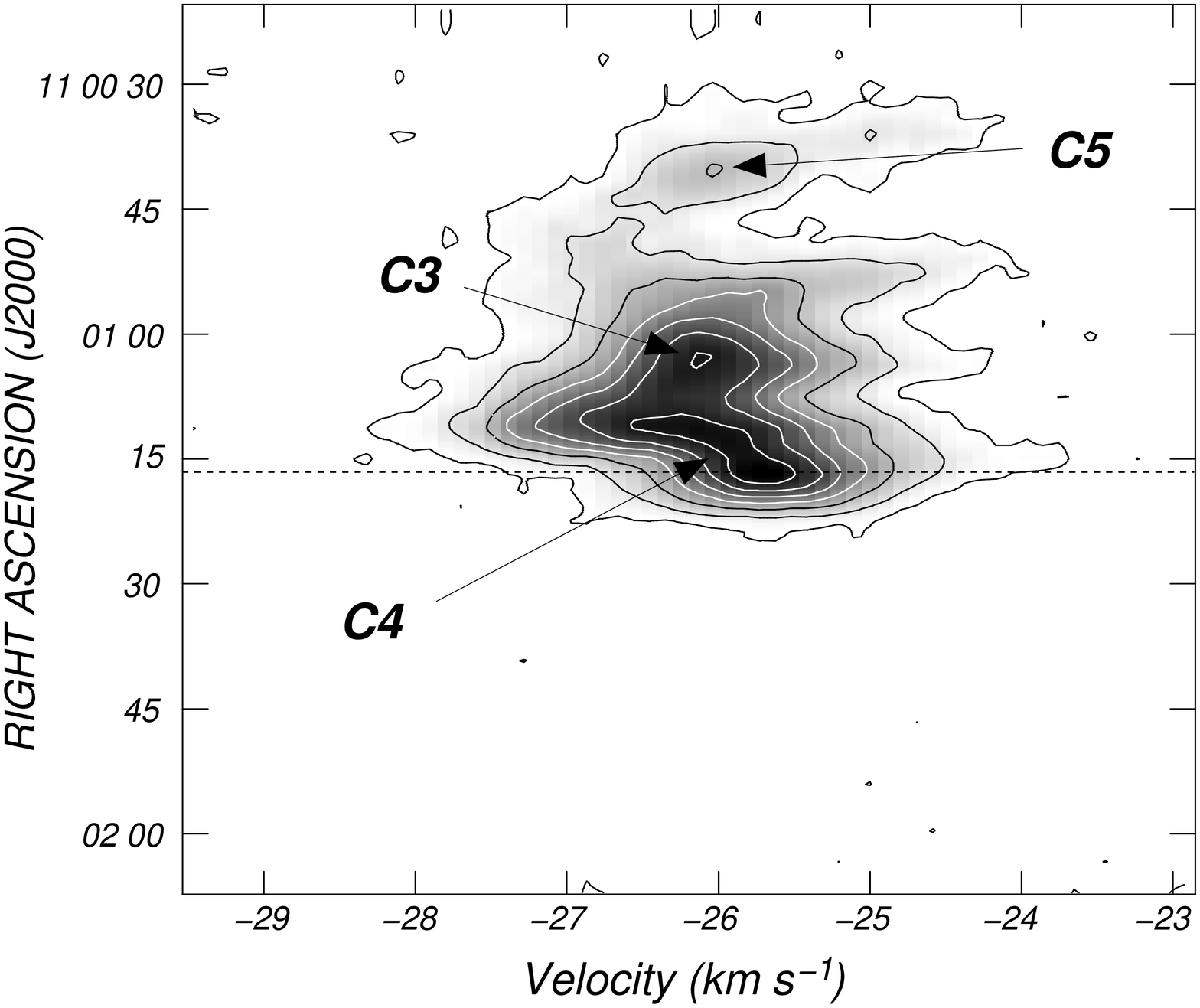}
\caption{ $^{13}$CO position-velocity map along Dec.(J2000) = $-$59\gra51\arcmin20\arcsec. The position of the center of \ngc\ is indicated by the dotted line.   }
\label{fig:gradiente}
\end{figure}

 In the previous section, we analyzed the line width of the  averaged clump spectra to look for signposts of a kinematical disturbance produced by the \hii\ region. Different from  expected,  the averaged spectrum   of clump C4  does not seem to be significantly broadened (when compared with the rest of the clumps) by the action of the \hii\ region.  However, a careful inspection at Fig. \ref{fig:mosaico} shows that the spatial location of the  peak emission of C4 is slightly displaced from RA. = 11$^h$01$^m$14$^s$  to RA. = 11$^h$01$^m$16$^s$ in the velocity range from $-$27.8 to $-$24.9 \kms, which gives rise to a small velocity gradient. In  Fig.\ref{fig:gradiente} we show the position-velocity map along Dec.(J2000) = $-$59\gra51\arcmin20\arcsec, slicing clumps C3, C4, and C5.  Unlike C3 and C5, a noticeable velocity gradient in the peak emission   of C4 is detected  around the central position of the IR nebula. This velocity gradient can be directly connected to the nebula and  could be interpreted as a significant sign of the disturbance in the molecular gas next to the ionization front. According to this interpretation, the expansion of the  ionized gas might be  affecting the kinematics of the molecular gas adjacent to it (clump C4) giving rise to this kinematical feature. An alternative explanation might be provided from the analysis of the  870 $\mu$m emission (see Sect. 3.5). In  Figs.  \ref{fig-yso}  and  \ref{yso-hcn-870}    two molecular cores  (identified in Sect. 3.5 as D1 and D2) can be noticed in the 870 $\mu$m emission (HPBW = 19$''$) inside the  \coc\ and  \cod\ emission of clump C4. Probably, C4 is composed by two different molecular cores at slightly different velocities which are not resolved by  carbon monoxide observations (HPBW = 27$''$ - 28$''$). These cores can be also noticed  in the \hcn\ emission (see Figs. \ref{fig:hcn} and \ref{yso-hcn-870}).

\subsection{Denser gas: LVG modeling}

\begin{figure}[h!]
\centering
\includegraphics[width=250pt]{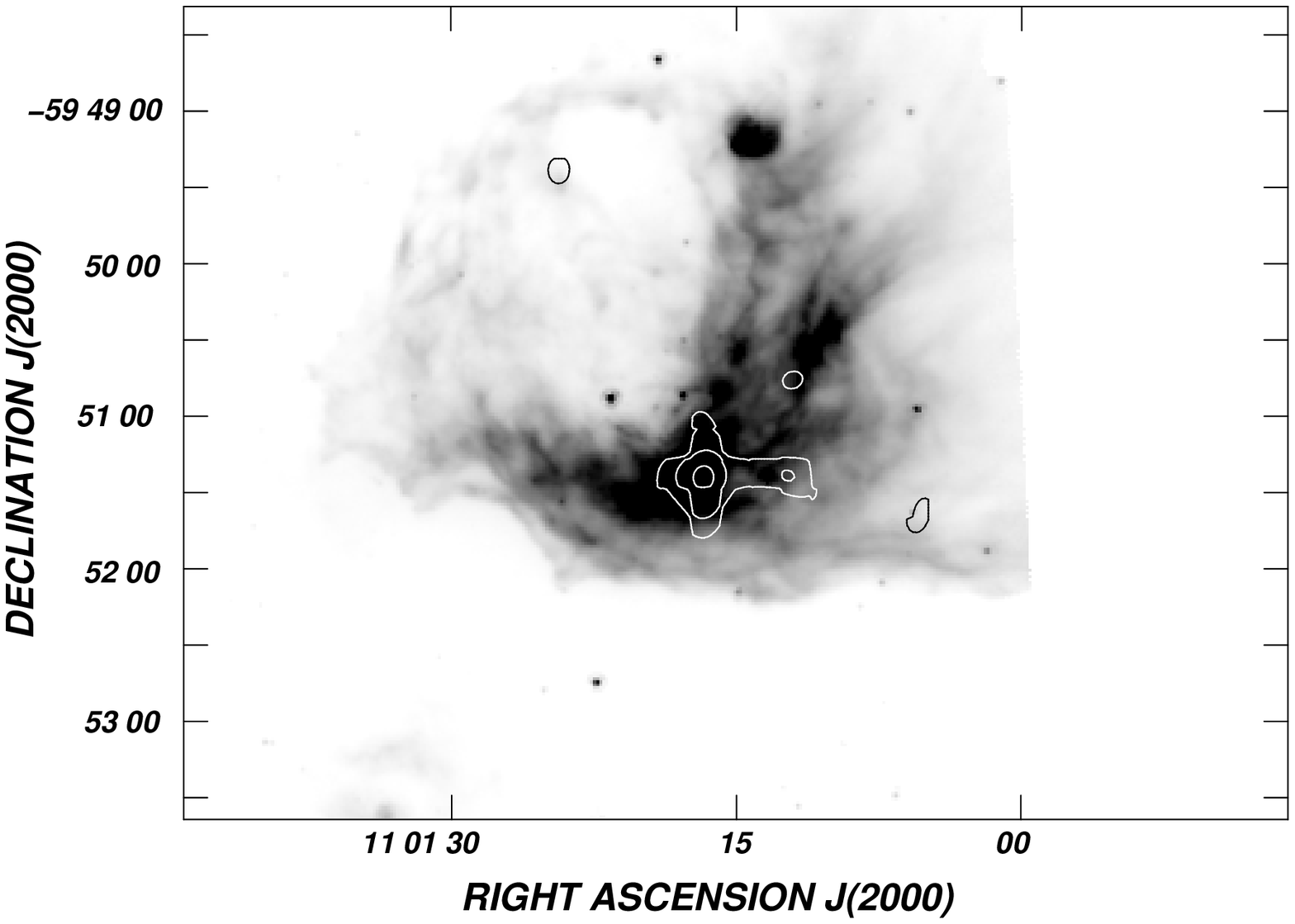}
\caption{ Mean HCN emission in the velocity range from $-26.5$ to $-25.5$ \kms (contours) superimposed onto the IRAC-4 image of the nebula (grays).  The lowest temperature contour is 0.55 K \kms\ ($\sim$5 rms) and  contour spacing temperature is 0.33  K \kms.  }
\label{fig:hcn}
\end{figure}

 In Fig.  \ref{12-13-18} we have shown that clump C4 has the only detection in the HCN line, at a velocity of $\sim$ $-$26 \kms. In Fig. \ref{fig:hcn} we show  an overlay of the mean HCN emission in the velocity interval from $-$26.5 \kms to $-$25.5 \kms\ onto the IRAC-GLIMPSE 4 emission of the nebula. 
The image shows four small sources placed at  \radec\ = (11$^h$01$^m$05$^s$, $-$59\gra51\arcmin40\arcsec),   \radec\ = (11$^h$01$^m$25$^s$, $-$59\gra49\arcmin22\arcsec),   \radec\ = (11$^h$01$^m$12$^s$, $-$59\gra50\arcmin40.5\arcsec),  and     \radec\ = (11$^h$01$^m$14.6$^s$, $-$59\gra51\arcmin22.7\arcsec). The first three sources are observed in their detection limits ($\sim$ 5 rms) and will not be considered to further analysis. The fourth source is more extended and achieves a peak temperature of $\sim$2 K at  \radec\ = (11$^h$01$^m$16.7$^s$, $-$59\gra51\arcmin23\arcsec). 
\begin{figure*}
\centering
\includegraphics[width=370pt]{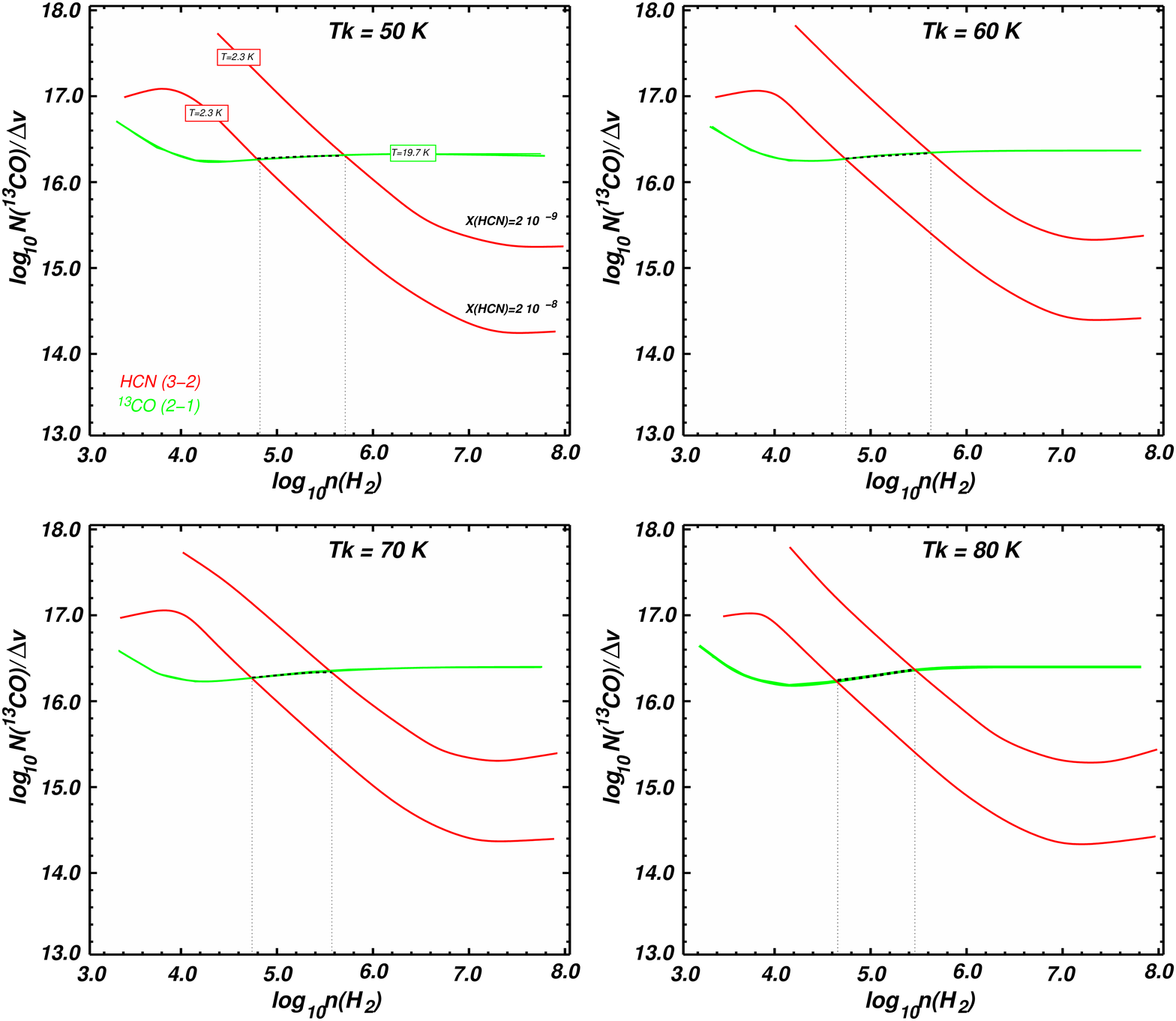}
\caption{LVG models for $^{13}$CO and HCN peak temperatures as function of normalized column density and volume density. We adopted abundances [$^{13}$CO]/[H$_2$] = 2 $\times$ 10$^{-6}$ and  [HCN]/[H$_2$] = (2 - 20) $\times$ 10$^{-8}$. The green lines trace the contours of  $^{13}$CO peak intensity, while the red lines trace the contours of HCN peak intensity. Dotted black lines are where solutions are coincident.      }
\label{fig:lvg}
\end{figure*}
A second and weaker peak temperature ($\sim$ 0.8 K) is observed at \radec\ = (11$^h$01$^m$11.6$^s$, $-$59\gra51\arcmin23.1\arcsec). The spatial location of this source is coincident with  the $^{13}$CO emission of C4, which clearly suggests that its emission represents the HCN counterpart of that molecular clump. The position of the two emission peaks in the HCN line at  \radec\ =  (11$^h$01$^m$16.7$^s$, $-$59\gra51\arcmin23\arcsec) and  \radec\ = (11$^h$01$^m$11.6$^s$, $-$59\gra51\arcmin23.1\arcsec) are almost coincident with the MSX C\hii\ region  candidate reported in Paper 1 as source 3 (see Figs. \ref{fig:hcn} and \ref{yso-hcn-870}).

Since the HCN molecule has a high critical density,   it has been suggested as ubiquitous high density molecular  gas tracer.  Since the spatial resolution of the HCN line is similar to that of the carbon monoxide lines, their emission can be used to obtain a robust estimation of the volume density using the large velocity gradient (LVG) formalism \citep{ss73,gk74} for radiative transfer of molecular emission lines.   We performed the LVG analysis with the code written by L. G. Mundy and implemented as part of the MIRIAD\footnote{http://www.cfa.harvard.edu/sma/miriad/packages/}  package of SMA. For a given kinetic temperature ($T_{\rm k}$), this program estimates the line radiation temperature  of a molecular transition as a function of the molecular column density (normalized by the line width) and H$_2$ volume density. Considering that $T_{\rm k}$ $\approx$ $T_{\rm exc}$ at    densities higher than 10$^{4}$ cm$^{-3}$ \citep{hay99}, and that $T_{\rm exc}$ = 57.2 K (see Table \ref{tabla:propiedades}) we adopted kinetic temperatures in the range  50 K $\leq$ $T_{\rm k}$ $\leq$ 80 K which are also typical temperatures derived for molecular clouds close to OB associations \citep{oh10}.  Nevertheless, we further find that  the derived densities are relatively insensitive to $T_{\rm k}$ in this range.  We use the $^{13}$CO molecule adopting a canonical abundance [$^{13}$CO]/[H$_2$] = 2 $\times$ 10$^{-6}$ \citep{d78}. For the case of HCN, its abundance is less certain making it the main source of error.    We adopted the abundance range  derived for  Orion-KL ([HCN]/[H$_2$] = 2 $\times$ 10$^{-8}$ - 2 $\times$ 10$^{-9}$; \citealt{sch92}).

In Fig. \ref{fig:lvg} we show the 50 $\times$ 50 model grids of \coc, and \hcn\ over a volume density range $n$(H$_2$) = 10$^3$ - 10$^8$ \cm3. For the \coc\ line, the normalized column density ranges are $N^{13}$(CO)/$\Delta v$ = 10$^{13}$ - 10$^{18}$ cm$^{-2}$ (\kms)$^{-1}$. At these high densities, the $^{13}$CO line is in collisional equilibrium and is almost independent of $n$(H$_2$). The HCN, however, has higher critical density and its line strength can be used to diagnose $n$(H$_2$) in this range.  The dotted line in each panel of Fig. \ref{fig:lvg} shows  the range of values where the solutions for both molecular species coincide, indicating that volume density in clump C4 is in the range 4 $\times$ 10$^{4}$ to 5.6 $\times$ 10$^{5}$ cm$^{-3}$.   These values are higher than that obtained from the LTE assumption ($n$(H$_2$)$_{\rm LTE}$ $\sim$ 1 $\times$ 10$^4$ cm$^{-3}$), although the latter is highly dependant on distance and geometry (assumed to be spherical) and a comparison might not be conclusive.          We have used in the LVG analysis HCN abundances obtained in Orion-KL, although lower abundances were derived for a number of galactic molecular clouds (e.g. 0.6 $\times$ 10$^{-10}$,  \citealt{joh03}; 7 $\times$ 10$^{-10}$,  \citealt{ten06}) which would highly shift the HCN lines up relative to $^{13}$CO, implying higher densities ($>$ 10$^{6}$ cm$^{-3}$). It is well accepted that volume densities higher than 10$^5$ cm$^{-3}$ are  critical for the initial condition of stellar formation \citep{elme02}. Further, two candidate YSOs were identified close to the HCN emission peak and projected onto the center of  C4 (see Fig. \ref{yso-hcn-870} in Sect. 3.6)    which suggests that star formation process may be occurring inside this dense clump. 
In order to obtain the $^{13}$CO column density, we multiplied the inferred value of $N^{13}$(CO)/$\Delta v$  by the line width, which yields to $N^{13}$(CO) $\sim$ 2.4 $\times$ 10$^{16}$ cm$^{-2}$.


\subsection{Continuum dust emission} 

Optically thin sub-millimeter continuum emission at 870 $\mu$m  is usually dominated by the thermal emission from cold dust, which is contained in dense material (e.g. dense molecular cores or filaments). 

\begin{table*}
\caption{Parameters of the dust clumps detected at 870 $\mu$m}
\begin{center}
\begin{tabular}{ c c c c c c c c c }
\hline
Dust clump &  RA   &   Dec.(J2000)  & $R_{\rm eff}$    &  $S_{870}$            &  $M_{\rm dust}$(20 K)$^{\ddag}$   &  $M_{\rm dust}$(30 K)$^{\ddag}$   &  $M_{\rm dust}$(40 K)$^{\ddag}$ & $^{13}$CO counterpart    \\
         &  ($^h$\ $^m$\ $^s$) & (\gra\ \arcmin\ \arcsec)      &    (arcsec)    &  (Jy)                & (\msun)              &   (\msun)             &    (\msun)         &        \\
\hline\hline
D1 \  & 11 : 01 : 17.4 &  $-$59 : 51 : 36  &     21        &  1.85 $\pm$ 0.22      &      1.59 $\pm$ 0.19     &   0.91 $\pm$ 0.11   &   0.63 $\pm$ 0.08      &   C4   \\
D2  \ & 11 : 01 : 12.6 &  $-$59 : 51 : 36   &     23        &   2.13 $\pm$ 0.26     &    1.84 $\pm$ 0.22    &  1.05 $\pm$ 0.13      &  0.73 $\pm$ 0.09          &   C4   \\
D3 \  &  11 : 01 : 05.2    &   $-$59 : 51 : 36     &     14        &    0.58 $\pm$ 0.09    &    0.50 $\pm$ 0.08   &  0.29 $\pm$ 0.04   &   0.20 $\pm$ 0.03   &   C3  \\
D4 \  & 11 : 01 : 08.7    &  $-$59 : 50 : 30    &     14        &       0.56 $\pm$ 0.10   &  0.48 $\pm$ 0.09  &  0.28 $\pm$ 0.05   &   0.19 $\pm$ 0.03   &   C7   \\
D5 \  &  11 : 01 : 13.5  &    $-$59 : 40 : 00     &  14  &   0.39 $\pm$ 0.10   &   2.59 $\pm$ 0.68$^{(\dag)}$ &  1.46 $\pm$ 0.37$^{(\dag)}$ & 1.02 $\pm$ 0.26$^{(\dag)}$    & C10 \\
D6 \  &  11 : 01 : 05.5    & $-$59 : 47 : 48   &      9       & 0.15 $\pm$ 0.04       &   1.00 $\pm$ 0.23$^{(\dag)}$  &  0.56 $\pm$ 0.15 $^{(\dag)}$ &  0.39 $\pm$ 0.10$^{(\dag)}$    &   C11      \\
\hline
\label{table:870}
\end{tabular}
\end{center}
{\scriptsize  Notes: $({\dag})$ Values obtained adopting a distance of $\sim$ 8 kpc. {\bf $(\ddag)$ Dust mass derived from the continuum emission at 870 $\mu$m.}}
\end{table*}

\begin{figure}[h!]
\centering
\includegraphics[width=250pt]{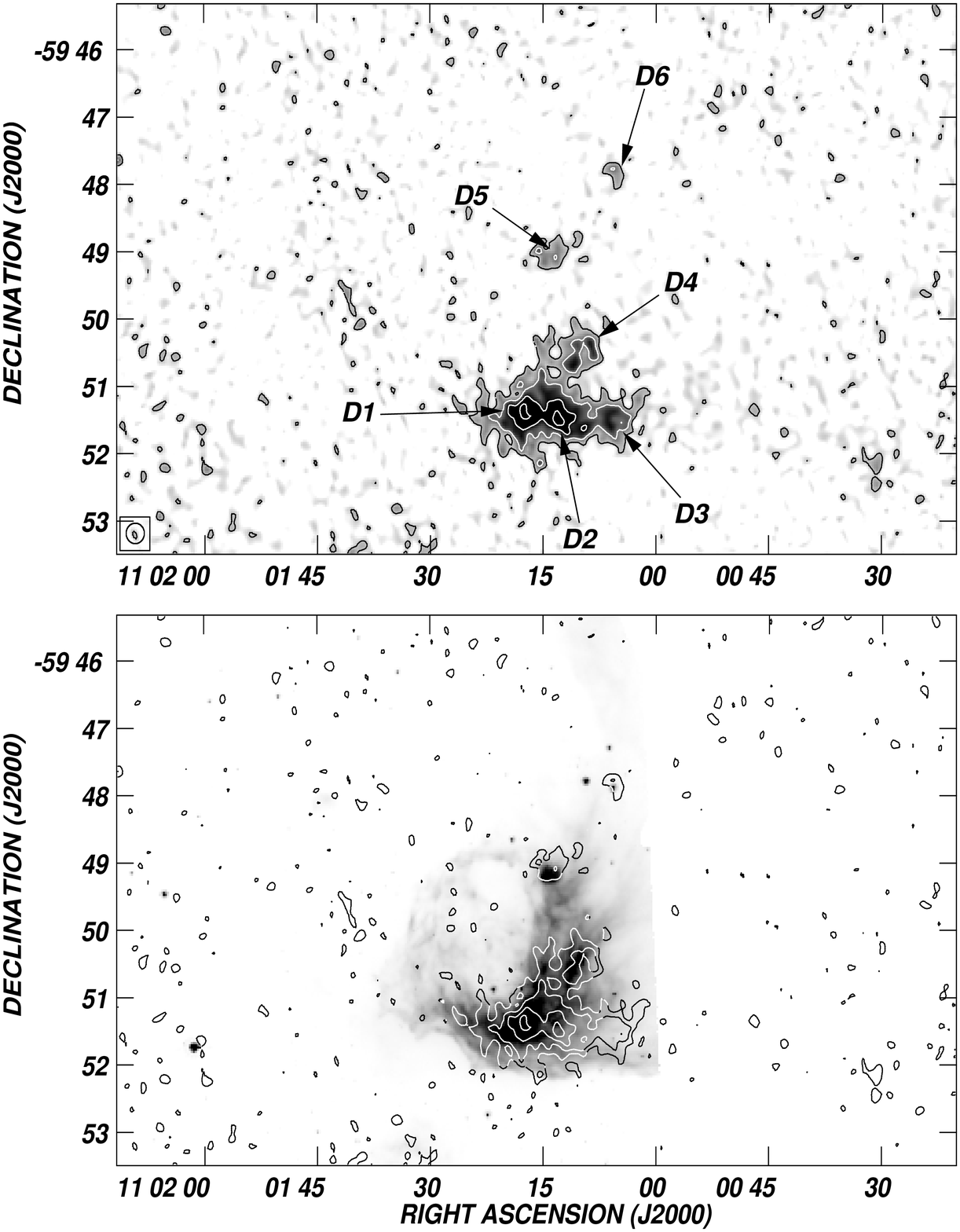}
\caption{{\it Upper panel:} 870 $\mu$m continuum emission map from ATLASGAL. The grayscale goes from 50 to 500 \mjyb. Contour levels correspond to 150 ($\sim$ 2.5 rms), 300, 500, and 700 \mjyb. {\it Bottom panel:} Overlay of IRAC image at 8 $\mu$m and the coutour levels of the upper panel. The grayscale goes from 10 to 180 MJy ster$^{-1}$. }
\label{870mi}
\end{figure}

In the upper panel of Fig. {\ref{870mi}} we display the image at 870 $\mu$m extracted from ATLASGAL. The image shows an extended source centered at \radec\ = (11$^h$01$^m$11$^s$, $-$59\gra51\arcmin). The brightest section of the source  extends along Dec.(J2000) = $-$59\gra51\arcmin36\arcsec, from RA. = 11$^h$01$^m$01$^s$  to 11$^h$01$^m$22$^s$. 
Three clumps can be easily identified, being  brighter the one to the east, and fainter the one to the west. Their positions are \radec\ = (11$^h$01$^m$17.4$^s$, $-$59\gra51\arcmin36\arcsec), \radec\ = (11$^h$01$^m$12.6$^s$, $-$59\gra51\arcmin36\arcsec),  and \radec\ = (11$^h$01$^m$05.2$^s$, $-$59\gra51\arcmin36\arcsec), indicated in the upper panel of Fig. \ref{870mi} as D1, D2, and D3, respectively. A fainter clump, named as D4, is placed at \radec\ = (11$^h$01$^m$08.7$^s$, $-$59\gra50\arcmin30\arcsec). The four dust clumps are immersed in a faint plateau of emission.

A comparison of the image at 870 $\mu$m and the $^{13}$CO images of Fig. 3 shows that D1 and D2 are the dust counterparts of C4, while D3 partially coincides with C3, and D4 with C7. Also note that the region of low molecular emission present at \radec\ = (11$^h$01$^m$06.3$^s$, $-$59\gra50\arcmin55\arcsec),  between C3 and C7, does not show emission at 870 $\mu$m. Clearly, the dust emission is the counterpart of the molecular emission. D1 and D2 are also detected in the HCN line (see Fig.\ref{yso-hcn-870} in Sect. 3.6).

Two additional patches of emission  are detected at 870 $\mu$m: one at \radec\ = (11$^h$01$^m$13.5$^s$, $-$59\gra49\arcmin00\arcsec) (D5) and the other at \radec\ = (11$^h$01$^m$05.5$^s$, $-$59\gra47\arcmin48\arcsec) (D6). Clumps D5 and D6 seem to  be the dust counterparts of the molecular  clumps C10 and C11, detected in the velocity interval $+$18.7 to $+$21.7 \kms\    (see Fig.  \ref{fig:mosaico}). In particular, D5 coincides with a faint source detected in the radio continuum at 4800 MHz and a bright source at 8 $\mu$m (see Fig. \ref{fig:ir-cont}). The 2MASS candidate young stellar objects $\#$1 and $\#$12  from Paper I  coincide with this region, suggesting that this is a candidate star forming region. Surprisingly, the emissions at both 870 $\mu$m and $^{13}$CO extend slightly to the north-west, opposite to the position of the central cavity of NGC\,3503 and the exciting stars, suggesting a relation to these object. However, as mentioned in Sect. 3.1,  the circular galactic rotation model predicts for a velocity of $\sim$ +20 \kms\ distances of about 8 kpc, far away from NGC 3503. As regards D6, it displays an arc-shaped morphology encircling  a point like source detected at 4.5 and 8 $\mu$m (labeled in Sect. 3.6 as 2MASS candidate YSO \#22 and WISE candidate YSO \#65).

Table \ref{table:870} lists flux densities and masses of the dust clumps.  Dust mass estimates  ($M_{\rm dust}$)   were obtained using Eq. \ref{eq:mdust} considering a conservative dust temperature range between 20 to 40 K. Values in the range 20 - 30 K are assumed for cold clumps and protostellar condensations  \citep{joh06,deh09}, while the last value  was derived from the emissions at 60 and 100 $\mu$m (see Paper I).     For dust clumps D5 and D6 we have adopted a distance of 8 kpc,   in common with molecular clumps C10 and C11, although a physical association with \ngc\  cannot be discarded without having information of the velocity of the ionized gas of MSX source \#1.

\subsection{A new identification of candidate YSOs }
     
As pointed out in Sect. 1, a search for candidate   YSOs  in the environs of \ngc\   was performed in Paper I using the IRAS, MSX, and 2MASS point source catalogs. To accomplish a more complete study, we have extended the search to a larger area around  the nebula and we have included new data. 
\begin{figure*}
\centering
\includegraphics[width=530pt]{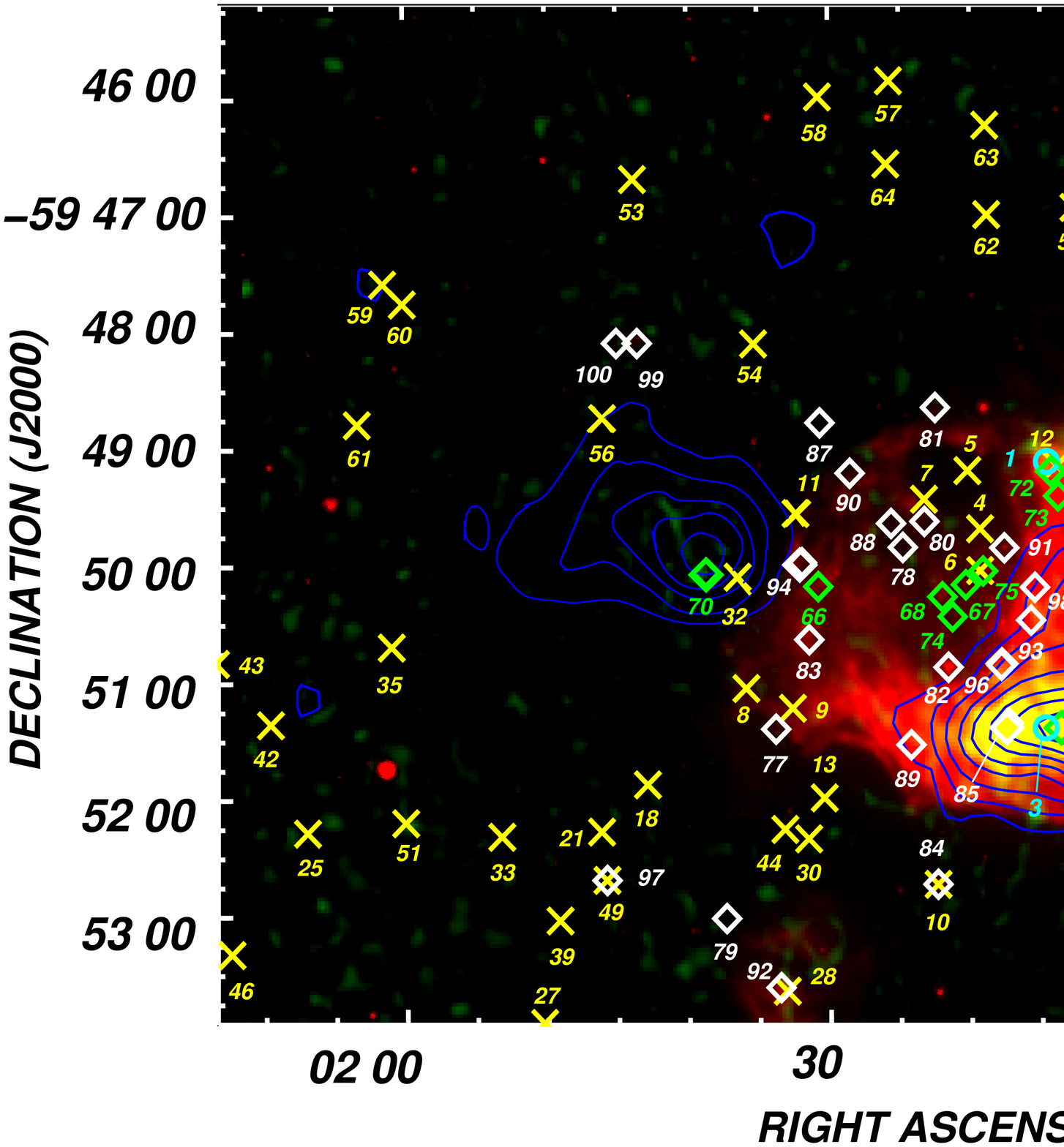}
\caption{ Composite image of \ngc\ and its environs. Red colour shows the IRAC-GLIMPSE emission at 8.0 $\mu$m, while  green/yellow tonalities show the ATLASGAL  870 $\mu$m continuum emission.  The \coc\ emission line integrated in the velocity ranges from  $-$27.8 to $-$23.7 \kms\   and $-$17.1 to $-$15.3 \kms  are  shown in blue contours. Thick light blue circles  indicate the position of the MSX  C\hii\ region candidates, while yellow crosses depict the position of the 2MASS sources with IR excess. Green  and white diamonds indicate the position of  WISE Class-I and Class-II candidates YSOs, respectively.  Numerical references of YSOs are based on Tables \ref{yso-2mass}  and \ref{yso-wise}. The size of the symbols do not  match the angular size of the sources.  }
\label{fig-yso}
\end{figure*}
We intend to study the star formation in the vicinity of \ngc\  by detecting all the candidate YSOs  (intrinsically reddened) and analyzing their position with respect to the  dust, ionized gas, and molecular gas.

In this new analysis we have used again the 2\,MASS catalog \citep{cu03}, which provides detections in $J$, $H$ and $K_s$ bands, to search for point sources with infrared excess. We have selected  only sources with signal-to-noise ratio (S/N) $>$ 10 (quality ``AAA'')  and followed the criteria of  \citet{co05}, to determine the parameter \hbox{$q$ = ($J$ - $H$) - 1.83 $\times$ ($H$ - $K_s$)}. Then, sources with $q <$ $-$0.15 (i.e. sources with IR  excess) were identified as candidate YSOs. In Table \ref{yso-2mass} we have listed the 2MASS candidate YSOs identified with the method explained above. For completeness, we have also included in the table  the three MSX C\hii region candidates reported in Paper I. For the sake of clarity, the numerical identification up to source \#12  is compatible to that of Paper I.  The spatial distribution of  2MASS and MSX sources is depicted in Fig. \ref{fig-yso}. As can be seen from this figure, from a total of 61 sources detected in the field, only 10 sources are projected onto the IR counterpart of \ngc\ or its central cavity, namely:  \#4, \#5, \#6, \#7, \#8, \#9, \#11, \#12, and \#22  (for  completeness, we have included in Fig. \ref{fig-yso} all the 2MASS candidate YSOs detected in the studied area). Sources \#4 to  \#12 were previously reported in Paper I, as possibly associated with the nebula.    Sources \#4, \#5, \#6, \#7 are projected towards the central cavity of \ngc, and lay inside  the radio continuum counterpart of the nebula (see Fig. \ref{fig:ir-cont}) while sources  \#8, \#9, and \#11  lay onto the IR arc  towards the eastern and northern sections of IR the nebula.  Low-intensity radio continuum emission is also seen projected onto these sources (see Fig. \ref{fig:ir-cont}). Source \#22 is coincident with a point-like source detected in the four IRAC bands, and correlates with  the emission at 870 $\mu$m of D6 (see Figs. \ref{870mi} and \ref{fig-yso}) and clump C11, suggesting that this candidate YSO is buried in  a ``cocoon'' of cold dust and dense gas. Its WISE counterpart can be classified as a Class-I candidate (source \#65; see below).  None of the above mentioned 2MASS sources are seen projected onto the molecular gas associated with the nebula at velocities between --27.8 to --23.7 \kms\ or --17.1 to --15.3 \kms  (see Fig. \ref{fig-yso}).    2MASS sources \#15, \#16, \#29, \#34, \#38,  \#40, \#45, and \#50)  appear projected onto the north-western and western  borders of the molecular gas in the velocity range from --27.8 to --23.7 \kms far from the nebula, thus, a physical association with the molecular environment of \ngc\ is uncertain.  Source \#32 is projected onto the molecular component at velocities between  --17.1 to --15.3 \kms (clump C9).

 Candidate YSOs are usually classified, according to their evolutionary phase, into two standard categories: 1) Class-I YSO which are  young protostellar sources embedded in dense infalling envelopes of gas and dust, and 2) Class-II YSO which are pre-main-sequence sources whose emission originates mainly in dense optically thick diks around the protostar (candidate T Tauri stars). These sources exhibit an infrared excess that  cannot be attributed to the ISM along the line of sight, but rather to the envelope and/or the disk surrounding the protostar. Class-III sources are usually referred to as pre-main-sequence (or main-sequence) field stars.  With the aim of identifying new candidate YSOs associated with \ngc , we have included in our analysis photometric data from the {\it Wide-field Infrared Survey Explorer} (WISE; \citealt{wr10}) obtained from the IPAC database\footnote{http://irsa.ipac.caltech.edu/cgi-bin/Gator/nph-dd}.  This survey maps the whole sky in four bands centered at 3.4, 4.6, 12, and 22 $\mu$m.  We have used the criteria  of  \citet{koe12},   biased   against the youngest and least massive candidate YSOs,  as follows:
\begin{table*}
\begin{center}
\caption{Candidate YSOs obtained from the  2\,MASS and MSX point source catalogs.}
\label{yso-2mass}
\begin{tabular}{ccccccc}
\hline
Source         &                \radec                   & MSX source &     F$_8$  &    F$_{12}$    & F$_{14}$   & F$_{21}$    \\
  No                & ($^h$ $^m$ $^s$, \gra \arcmin \arcsec)   &            &    (Jy)   &    (Jy)       & (Jy)      & (Jy) \\ 
\hline
     1	& 11 01 10.77, --59 49 13.9 & G289.4859+00.1420 & 1.49250 & 1.8145 & 0.9005 & 0.3180 \\
     2	& 11 01 11.18, --59 50 34.3 & G289.4993+00.1231 & 0.82699 & 1.4887 & 1.2237 & 4.3818\\
     3	& 11 01 13.57, --59 51 21.9 & G289.5051+00.1161 & 0.87911 & 2.8632 & 1.6759 & 5.6330\\

\hline
     &  \radec                                  &  2MASS source  & $J$   & $H$   & $K$   & \\
 $\#$     &  ($^h$ $^m$ $^s$, \gra \arcmin \arcsec)   &                & (mag) & (mag) & (mag) & \\ 
\hline
     4	& 11 01 19.34,  -59 49 42.2 & 11011934-5949422 & 13.277 & 13.047 & 12.806 & \\
     5	& 11 01 20.21,  -59 49 12.3 & 11012021-5949123 & 15.084 & 14.837 & 14.506 & \\
     6	& 11 01 19.33,  -59 50 03.7 & 11011933-5950037 & 14.539 & 14.100 & 13.696 & \\
     7	& 11 01 23.29,  -59 49 26.9 & 11012329-5949269 & 14.862 & 14.380 & 14.027 & \\
     8	& 11 01 35.91,  -59 51 03.7 & 11013591-5951037 & 15.237 & 14.592 & 14.106 & \\
     9	& 11 01 32.63,  -59 51 14.0 & 11013263-5951140 & 15.633 & 14.946 & 14.404 & \\
    10	& 11 01 22.38,  -59 52 44.9 & 11012238-5952449 & 12.747 & 11.744 & 10.836 & \\
    11	& 11 01 32.38,  -59 49 33.9 & 11013238-5949339 & 14.754 & 14.243 & 13.873 & \\
    12	& 11 01 13.90,  -59 49 09.9 & 11011390-5949099 & 14.888 & 13.661 & 12.827 & \\
    13	& 11 01 30.41,  -59 52 00.2 & 11013041-5952001 & 13.110 & 13.006 & 12.857 &  \\
    14	& 11 00 35.64,  -59 53 14.1 & 11003564-5953141 & 15.385 & 14.976 & 14.512 &  \\
    15	& 11 00 38.07,  -59 46 06.3 & 11003807-5946063 & 14.239 & 13.649 & 13.229 &  \\
    16	& 11 00 41.91,  -59 45 53.5 & 11004191-5945535 & 15.421 & 14.638 & 14.128 &  \\
    17	& 11 00 32.05,  -59 50 12.6 & 11003205-5950126 & 14.676 & 14.123 & 13.693 & \\
    18	& 11 01 42.92,  -59 51 52.9 & 11014292-5951529 & 15.091 & 14.775 & 14.497  &\\
    19	& 11 00 35.99,  -59 45 56.6 & 11003599-5945566 & 15.347 & 14.712 & 14.238 &\\
    20	& 11 00 57.48,  -59 52 32.2 & 11005748-5952322 & 15.124 & 14.855 & 14.472 &\\
    21	& 11 01 46.19,  -59 52 17.1 & 11014619-5952171 & 15.145 & 14.657 & 14.284 &\\
    22	& 11 01 05.71,  -59 47 53.4 & 11010571-5947534 & 15.142 & 12.941 & 11.305 &\\
    23	& 11 00 56.09,  -59 47 03.5 & 11005609-5947035 & 15.796 & 15.042 & 14.527 &\\
    24	& 11 00 35.10,  -59 48 39.7 & 11003510-5948397 & 14.950 & 14.499 & 14.114 &\\
    25	& 11 02 07.01,  -59 52 17.1 & 11020701-5952170 & 12.534 & 12.421 & 12.269 &\\
    26	& 11 00 35.21,  -59 48 43.5 & 11003521-5948435 & 15.413 & 14.571 & 14.017 &\\
    27	& 11 01 50.30,  -59 53 55.2 & 11015030-5953552 & 13.501 & 13.435 & 13.264 &\\
    28	& 11 01 33.09,  -59 53 39.4 & 11013309-5953394 & 15.326 & 14.602 & 14.097 &\\
    29	& 11 00 45.46,  -59 48 42.5 & 11004546-5948425 & 14.941 & 14.390 & 13.905 &\\
    30	& 11 01 31.53,  -59 52 21.1 & 11013153-5952211 & 14.358 & 14.177 & 13.985 &\\
    31	& 11 00 26.52,  -59 49 11.2 & 11002652-5949111 & 15.785 & 15.019 & 14.418 &\\
    32	& 11 01 36.49,  -59 50 06.5 & 11013649-5950065 & 15.528 & 14.928 & 14.352 &\\
    33	& 11 01 53.25,  -59 52 19.4 & 11015325-5952194 & 14.362 & 14.091 & 13.856 &\\
    34	& 11 00 57.88,  -59 48 53.6 & 11005788-5948536 & 15.784 & 15.092 & 14.447 &\\
    35	& 11 02 00.99,  -59 50 41.7 & 11020099-5950417 & 15.593 & 14.831 & 14.246 &\\
    36	& 11 00 31.45,  -59 46 49.3 & 11003145-5946493 & 14.796 & 13.869 & 13.266 &\\
    37	& 11 01 01.56,  -59 52 41.3 & 11010156-5952413 & 15.020 & 14.670 & 14.388 &\\
    38	& 11 00 46.66,  -59 52 03.7 & 11004666-5952037 & 15.245 & 14.724 & 14.246 &\\
    39	& 11 01 49.19,  -59 53 02.6 & 11014916-5953026 & 14.233 & 14.117 & 13.899 &\\
    40	& 11 00 46.77,  -59 49 08.3 & 11004677-5949083 & 14.713 & 14.509 & 14.121 &\\
    41	& 11 00 47.55,  -59 52 53.9 & 11004755-5952539 & 15.062 & 14.778 & 14.451 &\\
    42	& 11 02 09.58,  -59 51 21.2 & 11020958-5951212 & 14.368 & 13.942 & 13.624 &\\
    43	& 11 02 12.42,  -59 53 19.2 & 11021242-5953192 & 14.666 & 14.477 & 14.227 &\\
    44	& 11 01 33.21,  -59 52 16.2 & 11013321-5952162 & 14.667 & 14.421 & 14.131 &\\
    45	& 11 00 53.27,  -59 50 14.1 & 11005327-5950141 & 15.318 & 14.585 & 14.094 &\\
    46	& 11 02 13.44,  -59 50 49.5 & 11021344-5950495 & 14.549 & 14.443 & 14.244 &\\
    47	& 11 00 32.75,  -59 45 52.0 & 11003275-5945520 & 14.577 & 14.018 & 13.628 &\\
    48	& 11 00 45.93,  -59 53 36.3 & 11004593-5953363 & 13.534 & 13.232 & 12.983 &\\
    49	& 11 01 45.83,  -59 52 41.9 & 11014583-5952419 & 14.245 & 13.271 & 12.580 &\\
    50	& 11 00 40.21,  -59 46 50.0 & 11004021-5946500 & 10.339 & 10.258 & 10.085 &\\
    51	& 11 02 00.06,  -59 52 12.1 & 11020005-5952121 & 15.819 & 14.957 & 14.206 &\\
    52	& 11 00 29.66,  -59 46 05.5 & 11002966-5946055 & 13.880 & 12.898 & 12.177 &\\
    53	& 11 01 43.81,  -59 46 41.8 & 11014381-5946418 & 15.025 & 14.770 & 14.478 &\\
    54	& 11 01 35.32,  -59 48 06.6 & 11013532-5948066 & 14.097 & 13.781 & 13.520 &\\
    55	& 11 01 12.65,  -59 46 57.3 & 11011265-5946573 & 14.795 & 14.540 & 14.239 &\\
    56	& 11 01 46.02,  -59 48 44.6 & 11014602-5948446 & 13.965 & 13.863 & 13.653 &\\
    57	& 11 01 25.70,  -59 45 51.8 & 11012570-5945518 & 14.413 & 14.381 & 14.241 &\\
    58	& 11 01 30.72,  -59 45 59.9 & 11013072-5945599 & 13.575 & 13.327 & 13.071 &\\
    59	& 11 02 01.50,  -59 47 35.2 & 11020150-5947352 & 13.610 & 13.485 & 13.333 &\\
    60	& 11 02 00.12,  -59 47 45.5 & 11020012-5947455 & 14.893 & 14.482 & 14.131 &\\
    61	& 11 02 03.37,  -59 48 47.4 & 11020336-5948474 & 15.734 & 15.127 & 14.512 &\\
    62	& 11 01 18.81,  -59 47 00.8 & 11011881-5947008 & 15.467 & 14.630 & 14.039 &\\
    63	& 11 01 18.92,  -59 46 14.8 & 11011892-5946148 & 15.156 & 14.839 & 14.500 &\\
    64	& 11 01 25.91,  -59 46 34.5 & 11012591-5946345 & 14.337 & 14.136 & 13.938 &\\

\hline 
\end{tabular}
\end{center}
\end{table*}   
\begin{table*}
\begin{center}
\caption{Candidate YSOs obtained from the WISE point source catalog. }
\label{yso-wise}
\begin{tabular}{ccccccccc}
\hline
  Source  & Class     & \radec                               & WISE source & [3.4]  & [4.6]  & [12.0]  & [22.0] & Matching with    \\
    No               &           & ( $^h$ $^m$ $^s$, \gra \arcmin \arcsec ) &             &  (mag) & (mag)  & (mag)  & (mag)&  MSX and  \\
 &&&&&&&& 2MASS sources\\
\hline
    65 &  I 	 & 11 01 05.73,  -59 47 53.5 & J110105.73-594753.5 & 9.574 & 8.360 & 5.252 & 2.904   & $\#$ 22  \\
    66 &  I	 & 11 01 30.79,  -59 50 11.6 & J110130.79-595011.6 & 12.473 & 11.446 & 6.191 & 5.443 &   \\
    67 &  I	 & 11 01 20.38,  -59 50 10.9 & J110120.35-595010.9 & 12.500 & 11.435 & 6.332 & 3.443 &   \\
    68 &  I	 & 11 01 22.08,  -59 50 17.1 & J110122.01-595017.1 & 13.019 & 11.657 & 6.277 & 2.017 &  \\
    69 &  I	 & 11 01 10.86,  -59 48 57.3 & J110110.86-594857.3 & 11.708 & 10.446 & 5.400 & 2.790 &  \\
    70 &  I	 & 11 01 38.74,  -59 50 04.4 & J110138.74-595004.4 & 13.295 & 11.985 & 8.423 & 4.506 &  \\
    71 &  I	 & 11 01 13.50,  -59 51 23.8 & J110113.50-595123.8 & 8.839 & 7.820 & 3.518 & 0.195   & $\sim$  $\#$ 3 \\
    72 &  I	 & 11 01 14.33,  -59 49 12.8 & J110114.33-594912.8 & 9.465 & 8.460 & 3.535 & 0.440   & $\#$ 12 and $\#$ 1  \\
    73 &  I	 & 11 01 13.81,  -59 49 25.5 & J110113.81-594925.5 & 10.826 & 9.792 & 5.056 & 5.581  &  \\
    74 &  I	 & 11 01 21.30,  -59 50 27.3 & J110121.30-595027.3 & 13.165 & 11.902 & 7.195 & 4.070 &  \\
    75 &  I	 & 11 01 19.13,  -59 50 03.9 & J110119.13-595003.9 & 12.388 & 11.244 & 6.155 & 2.784 & $\#$ 6 \\
    76 &  I	 & 11 01 05.73,  -59 51 39.2 & J110105.73-595139.2 & 11.520 & 10.477 & 5.249 & 2.930 &  \\
\hline    
\hline
  source  & Class   & \radec                               & WISE source & [3.4]  & [4.6]  & [12.0]  & [22.0] &   \\
    $\#$  &        & ( $^h$ $^m$ $^s$, \gra \arcmin \arcsec )&             &  (mag) & (mag)  & (mag)  & (mag) &  \\  
\hline
     77 &  II	 & 11 01 33.82,  -59 51 24.7 & J110133.82-595124.7 & 11.083 & 10.696 & 6.622 & 3.169   &\\
     78 &  II	 & 11 01 24.80,  -59 49 51.6 & J110124.80-594951.6 & 13.012 & 12.209 & 7.309 & 3.537   &\\
     79 &  II	 & 11 01 37.36,  -59 53 01.9 & J110137.36-595301.9 & 13.755 & 13.277 & 8.518 & 6.706   &\\
     80 &  II	 & 11 01 23.27,  -59 49 38.3 & J110123.27-594938.3 & 12.510 & 11.852 & 7.009 & 3.912   &\\
     81 &  II	 & 11 01 22.49,  -59 48 39.8 & J110122.49-594839.8 & 12.270 & 11.805 & 7.058 & 3.972   &\\
     82 &  II	 & 11 01 21.60,  -59 50 53.3 & J110121.60-595053.3 & 8.906 & 8.003 & 5.084 & 1.147     &\\
     83 &  II	 & 11 01 31.44,  -59 50 38.8 & J110131.44-595038.8 & 11.325 & 10.853 & 6.003 & 3.915   & \\
     84 &  II	 & 11 01 22.38,  -59 52 44.8 & J110122.38-595244.8 & 9.631 & 9.046 & 5.964 & 3.402     & $\#$ 10 \\
     85 &  II	 & 11 01 17.35,  -59 51 22.1 & J110117.35-595122.1 & 9.120 & 8.162 & 3.628 & 1.149     &\\
     86 &  II	 & 11 00 29.67,  -59 46 05.5 & J110029.67-594605.5 & 11.125 & 10.635 & 8.156 & 5.558   &  $\#$ 52  \\
     87 &  II	 & 11 01 30.65,  -59 48 47.4 & J110130.65-594847.4 & 13.315 & 12.992 & 10.876 & 6.954  & \\
     88 &  II	 & 11 01 25.62,  -59 49 39.3 & J110125.62-594939.3 & 11.862 & 11.122 & 7.408 & 4.760   & \\
     89 &  II	 & 11 01 24.25,  -59 51 33.1 & J110124.25-595133.1 & 9.862 & 9.281 & 4.447 & 4.481     &\\
     90 &  II	 & 11 01 28.54,  -59 49 13.4 & J110128.54-594913.4 & 11.943 & 11.598 & 6.835 & 4.735   &\\
     91 &  II	 & 11 01 17.62,  -59 49 52.0 & J110117.62-594952.0 & 10.820 & 10.080 & 6.688 & 1.612   &\\
     92 &  II	 & 11 01 33.53,  -59 53 37.5 & J110133.53-595337.5 & 11.280 & 10.437 & 5.819 & 3.931   & $\#$ 28  \\
     93 &  II	 & 11 01 15.75,  -59 50 29.3 & J110115.75-595029.3 & 9.352 & 8.891 & 4.124 & 1.614     &\\
     94 &  II	 & 11 01 31.93,  -59 49 59.2 & J110131.93-594959.2 & 10.784 & 10.340 & 6.082 & 4.162   &\\
     95 &  II	 & 11 01 11.41,  -59 49 09.6 & J110111.41-594909.6 & 10.341 & 9.843 & 5.005 & 2.870    &\\
     96 &  II	 & 11 01 17.84,  -59 50 52.4 & J110117.84-595052.4 & 8.748 & 8.193 & 5.036 & 5.552     &\\
     97 &  II	 & 11 01 45.82,  -59 52 42.0 & J110145.82-595242.0 & 11.722 & 11.118 & 9.209 & 6.565   & $\#$ 49  \\
     98 &  II	 & 11 01 15.46,  -59 50 12.4 & J110115.46-595012.4 & 9.949 & 9.430 & 4.714 & 1.120     &\\
     99 &  II	 & 11 01 43.54,  -59 48 06.1 & J110143.54-594806.1 & 12.745 & 11.920 & 7.143 & 5.265   &\\
     100 & II	 & 11 01 45.01,  -59 48 06.0 & J110145.01-594806.0 & 12.976 & 12.318 & 7.502 & 5.598   &\\
\hline 
\end{tabular}
\end{center}
\end{table*}
after removing contamination arising from background objects like galaxies (very red in [4.6]$-$[12]), broad-line active galactic nuclei (of similar colours as YSOs, but distinctly fainter) and resolved PAH emission regions (redder than the majority of YSOs),  we identified infrared excess sources demanding that 
\[ \begin{array}{ll}
[3.4] - [4.6] - \sigma_1 & > 0.25\\

[4.6] - [12.0] - \sigma_2  & > 1.0

\end{array} \]
where [3.4], [4.6], and [12.0] are the WISE bands 1, 2, and 3 magnitudes, respectively, and  $\sigma_1 = \sigma([3.4]-[4.6])$ and $\sigma_2 = \sigma([4.6]-[12.0])$ indicate the combined errors of  [3.4]$-$[4.6] and [4.6]$-$[12.0] colors, added in quadrature.  Class I sources are a sub-sample of this defined by 
\[ \begin{array}{ll}
[3.4] - [4.6] & > 1.0 \\

[4.6] - [12.0] & > 2.0

\end{array} \]
(the rest are Class II objects). For the method explained above, we have considered sources with error in magnitudes lower than 0.2 mag in bands 1, 2, and 3.
\begin{figure}
\centering
\includegraphics[width=240pt]{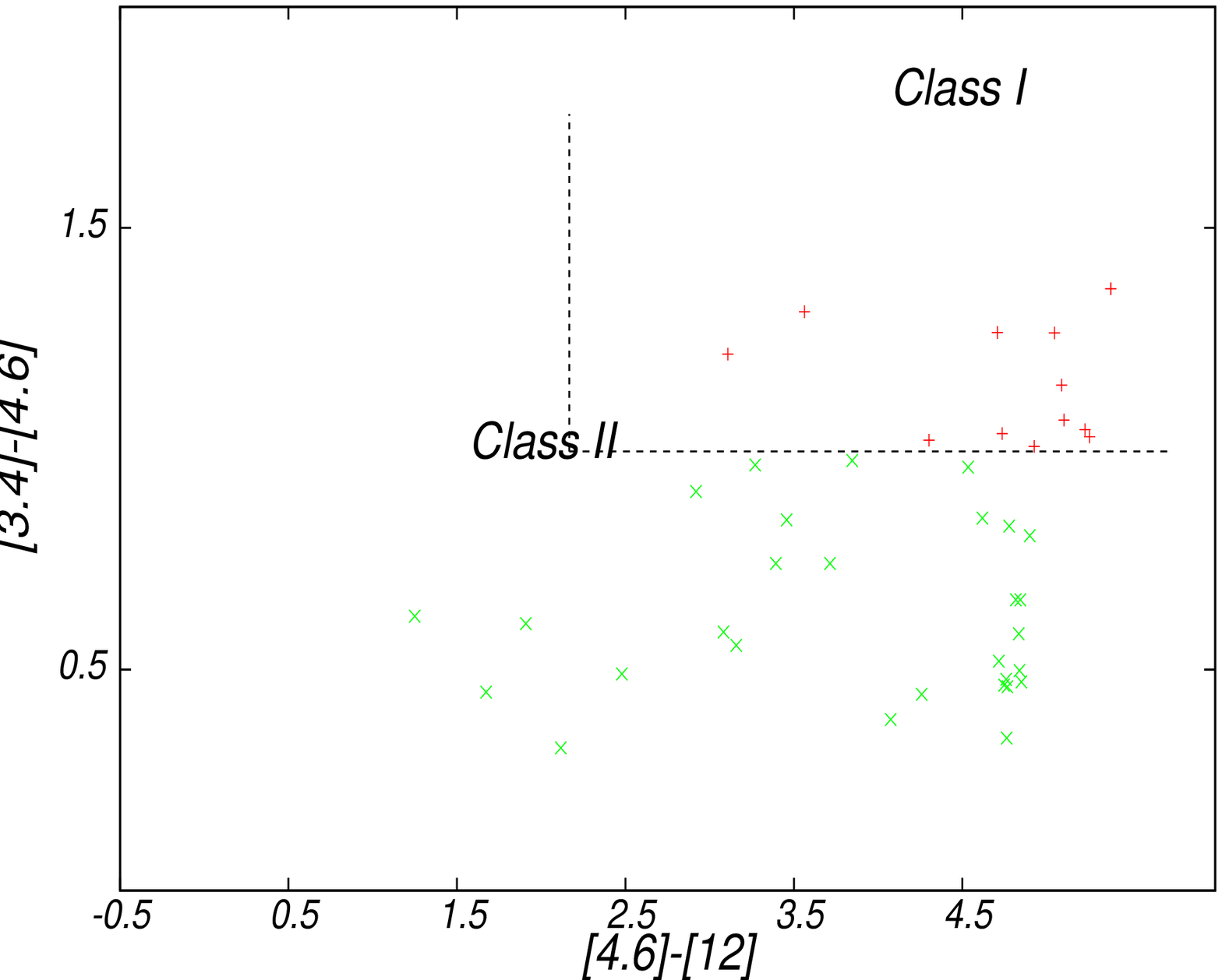}
\caption{ WISE band 1,2, and 3 color-color diagram showing the distribution of Class-I (in red) and Class-II (in green) candidate  YSOs.    }
\label{hist-cc}
\end{figure}

   In Table \ref{yso-wise} we list the candidate YSOs identified with the method explained above. For the sake of clarity, the numerical identification is succedent to that of Table \ref{yso-2mass}.    We found a total of 36 sources (12 Class-I and 24 Class-II candidates). In Fig.\ref{hist-cc}  we show the WISE band 1,2, and 3 color-color diagram  depicting the distribution of sources listed in Table  \ref{yso-wise}. The spatial location of Class-I and Class-II candidates is illustrated in  Fig. \ref{fig-yso}. Unlike 2MASS sources, the spatial distribution of WISE sources is rather  concentrated toward the nebula.  Sources  \#67, \#68, \#74, \#75 (which is coincident with 2MASS source \#6), \#78, \#80, \#88, and  \#91  are projected toward the center of the nebula and its radio continuum emission (see Fig.\ref{fig:ir-cont}), while sources \#66, \#77, \#81, \#83, \#90, and  \#94 are  seen projected onto the IR arc. Sources \#72 and \#73 are almost coincident with the MSX C\hii\ region candidate \#1 and 2MASS candidate YSO \#12.  Source \#70 is projected close to the peak of emission of the molecular component at velocities between --17.1 to --15.3 \kms\ (clump C9)

\begin{figure}
\centering
\includegraphics[width=260pt]{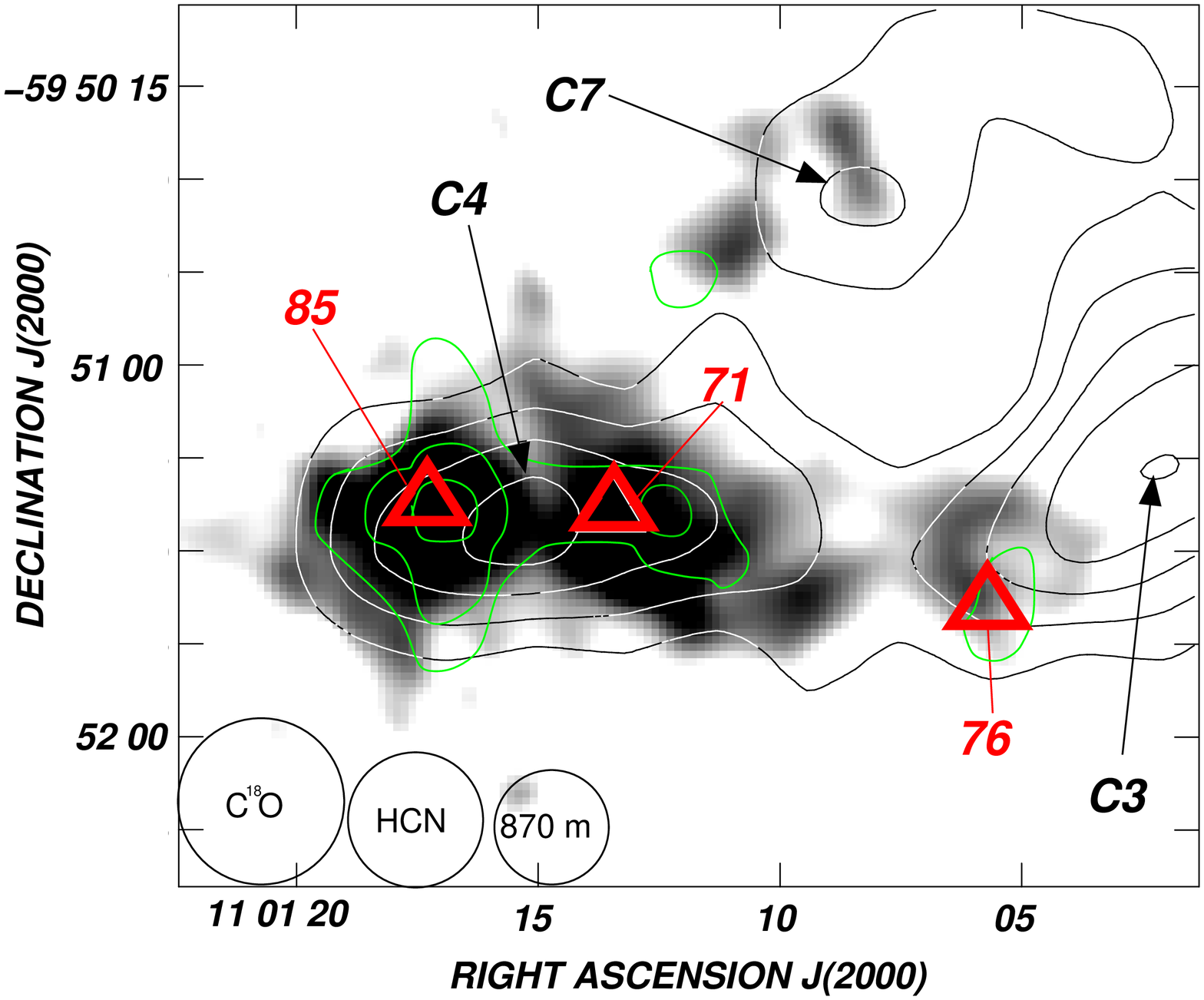}
\caption{Composite image showing the emission of the \hcn\ line (green contours), the \cod\ line (white/black contours), and the 870 $\mu$m continuum emission (grayscale)  in the central region of \ngc. The velocity ranges are $-$26.2 to $-$25.2 \kms\ and $-$26.5 to $-$25.5 \kms\ for the C$^{18}$O and HCN lines, respectively.  Red triangles depict the position of WISE candidate YSOs \#71, \#76, and  \#85.  The C$^{18}$O couterpart of clumps C3, C4, and C7 are indicated. The beam sizes are shown in the lower left corner.}  
\label{yso-hcn-870}
\end{figure}

None of the WISE candidate YSOs sources  mentioned so far are observed in the direction of any molecular clump detected in the velocity range from $-$27.8 to $-$23.7 \kms. Differently, three protostellar sources, \#71, \#76, and  \#85,  are observed in the direction of $^{13}$CO emission peaks corresponding to clumps C4 and C3, which are the densest clumps (see Sect. 3.2). In order to better trace the spatial distribution of the gas/dust and candidate YSOs in this region, we present  in Fig.\ref{yso-hcn-870} a composite image of HCN, C$^{18}$O, and 870 $\mu$m emission,   superimposed on the location of WISE sources \#71, \#76, and  \#85. Sources  \#71 and  \#85 are  projected onto dust clumps D2 and D1, respectively,  and outwardly a  C$^{18}$O peak emission of seen at \radec\ = (11$^h$01$^m$15$^s$, $-$59\gra51\arcmin20\arcsec) (C$^{18}$O counterpart of clump C4). Source  \#85 is almost projected onto the HCN peak at \radec\ = (11$^h$01$^m$16.7$^s$, $-$59\gra51\arcmin23\arcsec), while the position of source \#71 is close to the HCN peak seen at  \radec\ = (11$^h$01$^m$11.6$^s$, $-$59\gra51\arcmin23.1\arcsec) and the central position of the MSX C\hii\ region candidate \#3. This very likely suggests that these sources were formed inside high density molecular cores. As for source  \#76, this object is seen projected close to  the direction of  a  C$^{18}$O peak seen at  \radec\ = (11$^h$01$^m$2.5$^s$, $-$59\gra50\arcmin15\arcsec) (very likely the C$^{18}$O counterpart of clump C3) and coincident with the dust clump D3 seen at 870 $\mu$m. This protostellar source is also  projected onto the weaker HCN structure located at \radec\ = (11$^h$01$^m$05$^s$, $-$59\gra51\arcmin40\arcsec). As suggested in Sect. 3.2,  clump C4 is  likely exposed to extra compression, heating, and ionization from \ngc\ and \sfo,  which might have  triggered/enhanced the formation of protostellar candidates \#71 and  \#85 (and probably \#76)  in the densest regions of the molecular gas.  In that matter, it is also worth to stress the location of WISE sources   \#69, \#82, \#89, \#93, \#95,  \#96, and  \#98, which are spatially aligned perfectly following the eastern border of molecular clumps C4, C6, and C7. The location of these protostellar sources suggests that they are linked to the collected layers of molecular gas due to the expansion of the ionization front of  \ngc\ over its molecular environment, which    has probably  aided  the stellar formation activity along the external borders of the molecular gas.

To determine whether the fragmentation of the collected layer via ``collect and collapse'' process might have triggered the star formation in the environs of \ngc, in Paper I  we made use of the analytical  model of \citet{wi94} for expanding \hii\ regions. We obtained that $t_{\rm frag}$ $\sim$ 3.5 \x 10$^6$ yr and   $R_{\rm frag}$ $\sim$ 7.5 pc, which  are considerably larger than the age, $t_{\rm dyn}$,  and size, $R_{\rm  HII}$, of \ngc\   (derived assuming a classical expanding \hii\ region;  \citealt{dy97}). This lead us to  conclude that fragmentation of the collected layer triggered by the  expansion of the nebula is doubtful. We keep in mind, however, that newer  simulations have shown that  accretion and ionization may occur simultaneously for compact \hii\ regions, which makes their size unrelated on their age until late in their lifetimes \citep{pet10}.

Given the lack of certainty of the latter scenario, we have considered  alternative  approaches: probably, the formation of YSOs lying at the border of the \hii\ region (sources   \#69, \#82, \#89, \#93, \#95,  \#96, and  \#98) and inside clumps C4 and C3 (sources \#71,  \#76, and  \#85)   results from an interaction of the ionization front of \ngc\ (and \sfo) with pre-existing gravitationally bound molecular condensations (C3, C4, and C7; see Sect. 3.2), which has enhanced the formation of the protostellar objects (RDI process; \citealt{lela94}). Also, the formation of the protostellar candidates may be the result of  small-scale Jeans gravitational instabilities in the collected layers of molecular gas (e.g. \citealp{pom09,par11}). Higher spatial resolution molecular observations might help to shed some light on these issues.

\section{Summary}

Using  APEX \cob, \coc, \cod, and \hcn\ line data, and ATLASGAL 870 $\mu$m images, we carried out a multifrequency study of the molecular gas and dust associated with the \hii\ region/star forming region  \ngc. To analyze the star formation process in the region we  made use of the WISE and 2MASS data obtained from the IPAC archive. This work is a follow-up study of \citet{dvcca12}. The main results can be summarized as follows: 
\begin{enumerate}

\item The ionized gas of \ngc\     is expanding against the molecular gas component in the velocity range from $\sim$ $-$28 to $-$23 \kms. The morphology of the molecular gas close to  the nebula, the location of the PDR, and the  shape of radio continuum emission  confirm the ``champagne flow'' scenario proposed in Paper I.

\item New APEX observations allowed the small scale structure of the molecular gas associated with the nebula (previously reported in Paper I) to be fully imaged. We identified several  molecular clumps (C1 to C11) and studied their physical and dynamical properties to investigate the impact of  the expanding nebula and/or the southern bright rimmed cloud \sfo\ onto the molecular gas.  

\item We found that warmer clumps (C3, C4, and C7) are close to the \hii\ region, which is indicative of an  external heating source, most probably by photoionization of their  surface molecular layers by the intense UV field of Pis 17. Warmer clumps are also denser, which suggests that they are submitted to an external compression due to the expansion of \ngc. These clumps are likely molecular gas of the parental cloud that has been  swept up by the expansion of the ionization front and has been condensed. Clump  C4 is also adjacent to \sfo, which might explain its highest temperature and density. A noticeable velocity gradient is  detected in C4.  This gradient may  indicative of a kinematical disturbance, or else, be the consequence of two molecular cores at slightly different velocity that are not fully resolved in the carbon monoxide emissions.  

\item Clumps located near to \ngc\ also have the highest LTE and virialized masses. They also exhibit the highest $M_{\rm LTE}/M_{\rm vir}$\ ratio (predominantly $>$1), which,  according to the classical interpretation, indicates that they are gravitationally bound.  

\item  All molecular clumps exhibit line broadening  beyond the thermal width, which possibly indicates that  in some cases the line widths of the composite spectrum of the clumps have different velocities,  rather than showing turbulence effects. Also, spatially unresolved process like ouflows, infall, accretion, etc.  might be contributing in the broadening. 

\item We have analyzed the 870 $\mu$m emission, characteristic of filaments and dense molecular cores.  We detected emission only in the direction of clumps C3, C4, and C7. This is also indicative of high density gas.

\item We have presented some evidence of stellar formation in the region by detecting sources with IR excess. Unlike 2MASS candidate YSOs, WISE Class-I and Class-II candidates show a spatial distribution concentrated to the IR nebula. Several sources are detected along the external border of two of the densest molecular clumps (C4 and C7) which  suggests that they might be formed in  the compressed layers of molecular gas. Three sources are projected close to  HCN, C$^{18}$O, and 870 $\mu$m emission peaks (coincident with the position  of clumps C4 and C3). This very likely indicates that they were born inside high density  molecular cores, making clumps C4 and C3 excellent candidates to further investigate star formation with higher spatial resolution instruments like ALMA.  

\item Since the dynamical age and fragmentation time derived for  the molecular layer differ to the age and radius of the nebula (Paper I) we have excluded the ``collect and collapse'' scenario for the YSO formation. Instead, we have proposed here some alternative mechanisms, such as ``radiative-driven implosion''  in pre-existing gravitationally bound clumps (C3, C4, and C7), or small-scale Jeans gravitational instabilities in the swept-up layers of molecular gas.

\end{enumerate}

\begin{acknowledgements}

  We acknowledge the anonymous referee for his/her helpful comments that improved the presentation of this paper. This project was partially financed by CONICET of Argentina under projects PIP 112-800201-01299 and PIP 02488 and, UNLP under project 11/G120, and CONICYT Proyect PFB06. This research has made use of the NASA/ IPAC Infrared Science Archive, which is operated by the Jet Propulsion Laboratory, California Institute of Technology, under contract with the National Aeronautics and Space Administration. This work is based [in part] on observations made with the Spitzer Space Telescope, which is operated by the Jet Propulsion Laboratory, California Institute of Technology under a contract with NASA. This publication makes use of data products from the Two Micron All Sky Survey, which is a joint project of the University of Massachusetts and the Infrared Processing and Analysis Center/California Institute of Technology, funded by the National Aeronautics and Space Administration and the National Science Foundation. The MSX mission is sponsored by the Ballistic Missile Defense Organization (BMDO).

\end{acknowledgements}

\bibliographystyle{aa}
\bibliography{bibliografia-ngc3503}

\begin{thebibliography}{52}
\expandafter\ifx\csname natexlab\endcsname\relax\def\natexlab#1{#1}\fi

\bibitem[{{Allen}(1973)}]{allen73}
{Allen}, C.~W. 1973, { }, ed. {London: University of London, Athlone Press,
  c1973, 3rd ed.}

\bibitem[{{Benjamin} {et~al.}(2003){Benjamin}, {Churchwell}, {Babler}, {Bania},
  {Clemens}, {Cohen}, {Dickey}, {Indebetouw}, {Jackson}, {Kobulnicky},
  {Lazarian}, {Marston}, {Mathis}, {Meade}, {Seager}, {Stolovy}, {Watson},
  {Whitney}, {Wolff}, \& {Wolfire}}]{ben03}
{Benjamin}, R.~A., {Churchwell}, E., {Babler}, B.~L., {et~al.} 2003, \pasp,
  115, 953

\bibitem[{{Brand} \& {Blitz}(1993)}]{bb93}
{Brand}, J. \& {Blitz}, L. 1993, \aap, 275, 67

\bibitem[{{Cappa} {et~al.}(2009){Cappa}, {Rubio}, {Mart{\'{\i}}n}, \&
  {Romero}}]{ca09}
{Cappa}, C.~E., {Rubio}, M., {Mart{\'{\i}}n}, M.~C., \& {Romero}, G.~A. 2009,
  \aap, 508, 759

\bibitem[{{Comer{\'o}n} {et~al.}(2005){Comer{\'o}n}, {Schneider}, \&
  {Russeil}}]{co05}
{Comer{\'o}n}, F., {Schneider}, N., \& {Russeil}, D. 2005, \aap, 433, 955

\bibitem[{{Cutri} {et~al.}(2003){Cutri}, {Skrutskie}, {van Dyk}, {Beichman},
  {Carpenter}, {Chester}, {Cambresy}, {Evans}, {Fowler}, {Gizis}, {Howard},
  {Huchra}, {Jarrett}, {Kopan}, {Kirkpatrick}, {Light}, {Marsh}, {McCallon},
  {Schneider}, {Stiening}, {Sykes}, {Weinberg}, {Wheaton}, {Wheelock}, \&
  {Zacarias}}]{cu03}
{Cutri}, R.~M., {Skrutskie}, M.~F., {van Dyk}, S., {et~al.} 2003, VizieR Online
  Data Catalog, 2246, 0

\bibitem[{{Deharveng} {et~al.}(2010){Deharveng}, {Schuller}, {Anderson},
  {Zavagno}, {Wyrowski}, {Menten}, {Bronfman}, {Testi}, {Walmsley}, \&
  {Wienen}}]{deh10}
{Deharveng}, L., {Schuller}, F., {Anderson}, L.~D., {et~al.} 2010, \aap, 523,
  A6

\bibitem[{{Deharveng} {et~al.}(2012){Deharveng}, {Zavagno}, {Anderson},
  {Motte}, {Abergel}, {Andr{\'e}}, {Bontemps}, {Leleu}, {Roussel}, \&
  {Russeil}}]{deh12}
{Deharveng}, L., {Zavagno}, A., {Anderson}, L.~D., {et~al.} 2012, \aap, 546,
  A74

\bibitem[{{Deharveng} {et~al.}(2009){Deharveng}, {Zavagno}, {Schuller},
  {Caplan}, {Pomar{\`e}s}, \& {De Breuck}}]{deh09}
{Deharveng}, L., {Zavagno}, A., {Schuller}, F., {et~al.} 2009, \aap, 496, 177

\bibitem[{{Dickman}(1978)}]{d78}
{Dickman}, R.~L. 1978, \apjs, 37, 407

\bibitem[{{Dreyer} \& {Sinnott}(1988)}]{dr88}
{Dreyer}, J. L.~E. \& {Sinnott}, R.~W. 1988, {NGC 2000.0, The Complete New
  General Catalogue and Index Catalogue of Nebulae and Star Clusters by J.L.E.
  Dreyer and R. W. Sinnott }, ed. {}

\bibitem[{{Duronea} {et~al.}(2012){Duronea}, {Vasquez}, {Cappa}, {Corti}, \&
  {Arnal}}]{dvcca12}
{Duronea}, N.~U., {Vasquez}, J., {Cappa}, C.~E., {Corti}, M., \& {Arnal}, E.~M.
  2012, \aap, 537, A149

\bibitem[{{Dyson} \& {Williams}(1997)}]{dy97}
{Dyson}, J.~E. \& {Williams}, D.~A. 1997, {The physics of the interstellar
  medium}

\bibitem[{{Elmegreen}(2002)}]{elme02}
{Elmegreen}, B.~G. 2002, \apj, 577, 206

\bibitem[{{Elmegreen} \& {Lada}(1977)}]{elm77}
{Elmegreen}, B.~G. \& {Lada}, C.~J. 1977, \apj, 214, 725

\bibitem[{{Field} {et~al.}(2011){Field}, {Blackman}, \& {Keto}}]{fie11}
{Field}, G.~B., {Blackman}, E.~G., \& {Keto}, E.~R. 2011, \mnras, 416, 710

\bibitem[{{Georgelin} {et~al.}(2000){Georgelin}, {Russeil}, {Amram},
  {Georgelin}, {Marcelin}, {Parker}, \& {Viale}}]{g00}
{Georgelin}, Y.~M., {Russeil}, D., {Amram}, P., {et~al.} 2000, \aap, 357, 308

\bibitem[{{Goldreich} \& {Kwan}(1974)}]{gk74}
{Goldreich}, P. \& {Kwan}, J. 1974, \apj, 189, 441

\bibitem[{{G{\"u}sten} {et~al.}(2006){G{\"u}sten}, {Nyman}, {Schilke},
  {Menten}, {Cesarsky}, \& {Booth}}]{gu06}
{G{\"u}sten}, R., {Nyman}, L.~{\AA}., {Schilke}, P., {et~al.} 2006, \aap, 454,
  L13

\bibitem[{{Hayakawa} {et~al.}(1999){Hayakawa}, {Mizuno}, {Onishi}, {Yonekura},
  {Hara}, {Yamaguchi}, \& {Fukui}}]{hay99}
{Hayakawa}, T., {Mizuno}, A., {Onishi}, T., {et~al.} 1999, \pasj, 51, 919

\bibitem[{{Herbst}(1975)}]{H75}
{Herbst}, W. 1975, \aj, 80, 212

\bibitem[{{Heyer} {et~al.}(2009){Heyer}, {Krawczyk}, {Duval}, \&
  {Jackson}}]{hey09}
{Heyer}, M., {Krawczyk}, C., {Duval}, J., \& {Jackson}, J.~M. 2009, \apj, 699,
  1092

\bibitem[{{Hollenbach} \& {Tielens}(1997)}]{ht97}
{Hollenbach}, D.~J. \& {Tielens}, A.~G.~G.~M. 1997, \araa, 35, 179

\bibitem[{{Johnstone} {et~al.}(2003){Johnstone}, {Boonman}, \& {van
  Dishoeck}}]{joh03}
{Johnstone}, D., {Boonman}, A.~M.~S., \& {van Dishoeck}, E.~F. 2003, \aap, 412,
  157

\bibitem[{{Johnstone} {et~al.}(2006){Johnstone}, {Matthews}, \&
  {Mitchell}}]{joh06}
{Johnstone}, D., {Matthews}, H., \& {Mitchell}, G.~F. 2006, \apj, 639, 259

\bibitem[{{Koenig} {et~al.}(2012){Koenig}, {Leisawitz}, {Benford}, {Rebull},
  {Padgett}, \& {Assef}}]{koe12}
{Koenig}, X.~P., {Leisawitz}, D.~T., {Benford}, D.~J., {et~al.} 2012, \apj,
  744, 130

\bibitem[{{Langer} \& {Penzias}(1993)}]{lp93}
{Langer}, W.~D. \& {Penzias}, A.~A. 1993, \apj, 408, 539

\bibitem[{{Lefloch} \& {Lazareff}(1994)}]{lela94}
{Lefloch}, B. \& {Lazareff}, B. 1994, \aap, 289, 559

\bibitem[{{MacLaren} {et~al.}(1988){MacLaren}, {Richardson}, \&
  {Wolfendale}}]{ml88}
{MacLaren}, I., {Richardson}, K.~M., \& {Wolfendale}, A.~W. 1988, \apj, 333,
  821

\bibitem[{{Massi} {et~al.}(2007){Massi}, {de Luca}, {Elia}, {Giannini},
  {Lorenzetti}, \& {Nisini}}]{mas07}
{Massi}, F., {de Luca}, M., {Elia}, D., {et~al.} 2007, \aap, 466, 1013

\bibitem[{{Ohama} {et~al.}(2010){Ohama}, {Dawson}, {Furukawa}, {Kawamura},
  {Moribe}, {Yamamoto}, {Okuda}, {Mizuno}, {Onishi}, {Maezawa}, {Minamidani},
  {Mizuno}, \& {Fukui}}]{oh10}
{Ohama}, A., {Dawson}, J.~R., {Furukawa}, N., {et~al.} 2010, \apj, 709, 975

\bibitem[{{Ossenkopf} \& {Henning}(1994)}]{osse94}
{Ossenkopf}, V. \& {Henning}, T. 1994, \aap, 291, 943

\bibitem[{{Paron} {et~al.}(2011){Paron}, {Petriella}, \& {Ortega}}]{par11}
{Paron}, S., {Petriella}, A., \& {Ortega}, M.~E. 2011, \aap, 525, A132

\bibitem[{{Peters} {et~al.}(2010){Peters}, {Banerjee}, {Klessen}, {Mac Low},
  {Galv{\'a}n-Madrid}, \& {Keto}}]{pet10}
{Peters}, T., {Banerjee}, R., {Klessen}, R.~S., {et~al.} 2010, \apj, 711, 1017

\bibitem[{{Pinheiro} {et~al.}(2010){Pinheiro}, {Copetti}, \&
  {Oliveira}}]{pco10}
{Pinheiro}, M.~C., {Copetti}, M.~V.~F., \& {Oliveira}, V.~A. 2010, \aap, 521,
  A26+

\bibitem[{{Pomar{\`e}s} {et~al.}(2009){Pomar{\`e}s}, {Zavagno}, {Deharveng},
  {Cunningham}, {Jones}, {Kurtz}, {Russeil}, {Caplan}, \&
  {Comer{\'o}n}}]{pom09}
{Pomar{\`e}s}, M., {Zavagno}, A., {Deharveng}, L., {et~al.} 2009, \aap, 494,
  987

\bibitem[{{Rohlfs} \& {Wilson}(2004)}]{rw04}
{Rohlfs}, K. \& {Wilson}, T.~L. 2004, {Tools of Radioastronomy }, ed.
  {Springer-Verlag, Berlin-Heidelberg}

\bibitem[{{Romero} \& {Cappa}(2009)}]{ro09}
{Romero}, G.~A. \& {Cappa}, C.~E. 2009, \mnras, 395, 2095

\bibitem[{{Schilke} {et~al.}(1992){Schilke}, {Walmsley}, {Pineau Des Forets},
  {Roueff}, {Flower}, \& {Guilloteau}}]{sch92}
{Schilke}, P., {Walmsley}, C.~M., {Pineau Des Forets}, G., {et~al.} 1992, \aap,
  256, 595

\bibitem[{{Schuller}(2012)}]{sch12}
{Schuller}, F. 2012, in Society of Photo-Optical Instrumentation Engineers
  (SPIE) Conference Series, Vol. 8452, Society of Photo-Optical Instrumentation
  Engineers (SPIE) Conference Series

\bibitem[{{Schuller} {et~al.}(2009){Schuller}, {Menten}, {Contreras},
  {Wyrowski}, {Schilke}, {Bronfman}, {Henning}, {Walmsley}, {Beuther},
  {Bontemps}, {Cesaroni}, {Deharveng}, {Garay}, {Herpin}, {Lefloch}, {Linz},
  {Mardones}, {Minier}, {Molinari}, {Motte}, {Nyman}, {Reveret}, {Risacher},
  {Russeil}, {Schneider}, {Testi}, {Troost}, {Vasyunina}, {Wienen}, {Zavagno},
  {Kovacs}, {Kreysa}, {Siringo}, \& {Wei{\ss}}}]{sch09}
{Schuller}, F., {Menten}, K.~M., {Contreras}, Y., {et~al.} 2009, \aap, 504, 415

\bibitem[{{Scoville} \& {Solomon}(1973)}]{ss73}
{Scoville}, N.~Z. \& {Solomon}, P.~M. 1973, \apj, 180, 31

\bibitem[{{Siringo} {et~al.}(2007){Siringo}, {Weiss}, {Kreysa}, {Schuller},
  {Kovacs}, {Beelen}, {Esch}, {Gem{\"u}nd}, {Jethava}, {Lundershausen},
  {Menten}, {G{\"u}sten}, {Bertoldi}, {De Breuck}, {Nyman}, {Haller}, \&
  {Beeman}}]{sir07}
{Siringo}, G., {Weiss}, A., {Kreysa}, E., {et~al.} 2007, The Messenger, 129, 2

\bibitem[{{Sugitani} {et~al.}(1991){Sugitani}, {Fukui}, \& {Ogura}}]{so91}
{Sugitani}, K., {Fukui}, Y., \& {Ogura}, K. 1991, \apjs, 77, 59

\bibitem[{{Tennekes} {et~al.}(2006){Tennekes}, {Harju}, {Juvela}, \&
  {T{\'o}th}}]{ten06}
{Tennekes}, P.~P., {Harju}, J., {Juvela}, M., \& {T{\'o}th}, L.~V. 2006, \aap,
  456, 1037

\bibitem[{{Thompson} {et~al.}(2004){Thompson}, {Urquhart}, \& {White}}]{T04}
{Thompson}, M.~A., {Urquhart}, J.~S., \& {White}, G.~J. 2004, \aap, 415, 627

\bibitem[{{Urquhart} {et~al.}(2009){Urquhart}, {Morgan}, \& {Thompson}}]{U09}
{Urquhart}, J.~S., {Morgan}, L.~K., \& {Thompson}, M.~A. 2009, \aap, 497, 789

\bibitem[{{Vasquez} {et~al.}(2012){Vasquez}, {Rubio}, {Cappa}, \&
  {Duronea}}]{va12}
{Vasquez}, J., {Rubio}, M., {Cappa}, C.~E., \& {Duronea}, N.~U. 2012, \aap,
  545, A89

\bibitem[{{Vassilev} {et~al.}(2008){Vassilev}, {Meledin}, {Lapkin}, {Belitsky},
  {Nystr{\"o}m}, {Henke}, {Pavolotsky}, {Monje}, {Risacher}, {Olberg},
  {Strandberg}, {Sundin}, {Fredrixon}, {Ferm}, {Desmaris}, {Dochev},
  {Pantaleev}, {Bergman}, \& {Olofsson}}]{vas08}
{Vassilev}, V., {Meledin}, D., {Lapkin}, I., {et~al.} 2008, \aap, 490, 1157

\bibitem[{{Whitworth} {et~al.}(1994){Whitworth}, {Bhattal}, {Chapman},
  {Disney}, \& {Turner}}]{wi94}
{Whitworth}, A.~P., {Bhattal}, A.~S., {Chapman}, S.~J., {Disney}, M.~J., \&
  {Turner}, J.~A. 1994, \mnras, 268, 291

\bibitem[{{Wright} {et~al.}(2010){Wright}, {Eisenhardt}, {Mainzer}, {Ressler},
  {Cutri}, {Jarrett}, {Kirkpatrick}, {Padgett}, {McMillan}, {Skrutskie},
  {Stanford}, {Cohen}, {Walker}, {Mather}, {Leisawitz}, {Gautier}, {McLean},
  {Benford}, {Lonsdale}, {Blain}, {Mendez}, {Irace}, {Duval}, {Liu}, {Royer},
  {Heinrichsen}, {Howard}, {Shannon}, {Kendall}, {Walsh}, {Larsen}, {Cardon},
  {Schick}, {Schwalm}, {Abid}, {Fabinsky}, {Naes}, \& {Tsai}}]{wr10}
{Wright}, E.~L., {Eisenhardt}, P.~R.~M., {Mainzer}, A.~K., {et~al.} 2010, \aj,
  140, 1868

\bibitem[{{Yamaguchi} {et~al.}(1999){Yamaguchi}, {Saito}, {Mizuno}, {Mine},
  {Mizuno}, {Ogawa}, \& {Fukui}}]{Y99}
{Yamaguchi}, R., {Saito}, H., {Mizuno}, N., {et~al.} 1999, \pasj, 51, 791

\end{thebibliography}
 
\IfFileExists{\jobname.bbl}{}
{\typeout{}
\typeout{****************************************************}
\typeout{****************************************************}
\typeout{** Please run "bibtex \jobname" to optain}
\typeout{** the bibliography and then re-run LaTeX}
\typeout{** twice to fix the references!}
\typeout{****************************************************}
\typeout{****************************************************}
\typeout{}

}

\end{document}